\documentclass[sigconf]{acmart}

\AtBeginDocument{%
  }
\usepackage{enumitem}
\usepackage{xcolor}
\usepackage{graphicx} %

\newcommand{\revisemb}[1]{#1}
\newcommand{\revise}[1]{#1}
\newcommand{\revisel}[1]{#1}

\newcommand{\tool}{SketchDynamics}
\newcommand{\q}[1]{\textit{``#1''}}

\begin{document}

\title{\tool: Exploring Free-Form Sketches for Dynamic Intent Expression in Animation Generation
}

\author{Boyu Li}
\affiliation{%
  \department{Arts and Machine Creativity}
  \institution{
  The Hong Kong University of Science and Technology}
  \city{Hong Kong}
  \country{China}}
\email{blibr@connect.ust.hk}

\author{Lin-Ping Yuan}
\affiliation{%
\department{Computer Science and Engineering}
  \institution{The Hong Kong University of Science and Technology}
  \city{Hong Kong}
  \country{China}}
\email{yuanlp@cse.ust.hk}

\author{Zeyu Wang}
\affiliation{%
  \department{Computational Media and Arts}
  \institution{The Hong Kong University of Science and Technology (Guangzhou)}
  \city{Guangzhou}
  \country{China}}
\affiliation{%
  \department{Computer Science and Engineering}
  \institution{The Hong Kong University of Science and Technology}
  \city{Hong Kong}
  \country{China}}
\email{zeyuwang@ust.hk}

\author{Hongbo Fu}
\authornote{Corresponding author.}
\affiliation{%
\department{Arts and Machine Creativity}
  \institution{
  The Hong Kong University of Science and Technology}
  \city{Hong Kong}
  \country{China}}
\email{fuplus@gmail.com}

\copyrightyear{2026}
\acmYear{2026}
\setcopyright{cc}
\setcctype{by}
\acmConference[CHI '26]{Proceedings of the 2026 CHI Conference on Human Factors in Computing Systems}{April 13--17, 2026}{Barcelona, Spain}
\acmBooktitle{Proceedings of the 2026 CHI Conference on Human Factors in Computing Systems (CHI '26), April 13--17, 2026, Barcelona, Spain}
\acmPrice{}
\acmDOI{10.1145/3772318.3791071}
\acmISBN{979-8-4007-2278-3/2026/04}


\begin{abstract}
Sketching provides an intuitive way to convey dynamic intent in animation authoring (i.e., how elements change over time and space), making it a natural medium for automatic content creation. Yet existing approaches often constrain sketches to \revisel{fixed command tokens or predefined visual forms}, overlooking their free-form nature and the central role of humans in shaping intention. 
\revisel{To address this, we introduce an interaction paradigm where users convey dynamic intent to a vision–language model via free-form sketching, instantiated here in a sketch storyboard to motion graphics workflow. We implement an interface and improve it through a three-stage study with 24 participants}. The study shows how sketches convey motion with minimal input, how their inherent ambiguity requires users to be involved for clarification, and how sketches can visually guide video refinement. Our findings reveal the potential of sketch–AI interaction to bridge the gap between intention and outcome, and demonstrate its applicability to 3D animation and video generation.

\end{abstract}




  
\begin{teaserfigure}
  \includegraphics[width=0.9\textwidth]{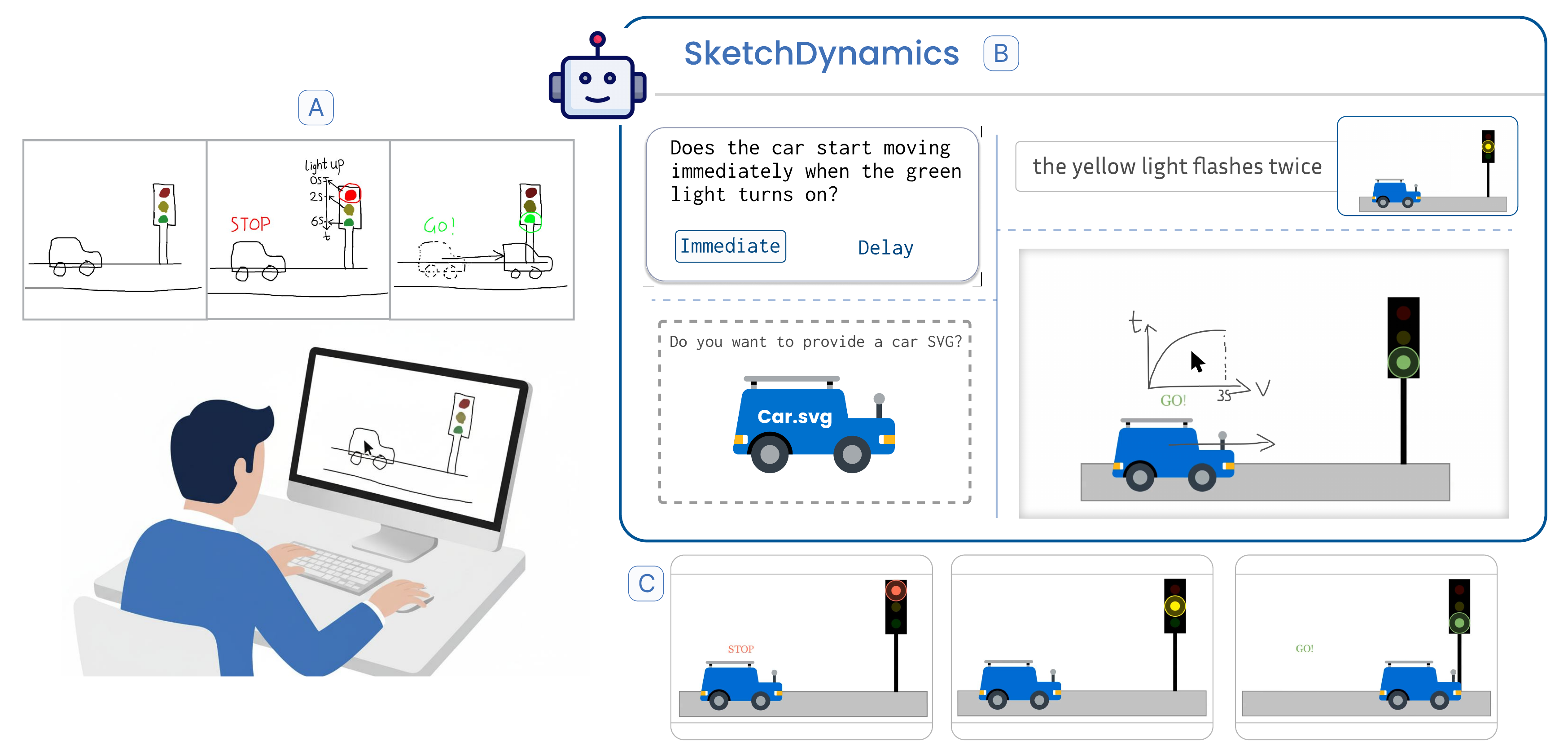}
  \centering
  \caption{Example interaction with \tool. (A) A user sketches a storyboard of a traffic-light scene (red stop, yellow flashing, green go). (B-Left) The system proactively raises clarification questions (e.g., whether the car starts moving immediately or after a delay, and whether a car SVG should be provided), and the user responds by selecting options or uploading assets. (B-Right) The user further refines the generated frames by specifying that the yellow light flashes twice and by sketching a velocity–time curve to indicate changes in the car’s speed. (C) At the bottom, the final video clips illustrate the produced animation sequence. }
  \label{fig:teaser}
  \Description{}
\end{teaserfigure}


\maketitle

\section{Introduction}

Sketching is one of the most accessible and intuitive ways for users to visually convey dynamic authoring intent, i.e., how they wish elements to change over time and space, particularly in expressive content domains such as film and animation~\cite{Chandrasegaran2018how,bhattacharjee2020survey}. Traditionally, sketches are used in two complementary ways to express dynamic intent. The first is as storyboards\revise{~\cite{griffith2024kato}}. From early Disney animation~\cite{madej2020disney} to contemporary film production~\cite{matthews2012sketch}, sketches are combined with scripts to depict keyframes and narrative flow. The second is as free-form annotations, where marks such as arrows, dashed lines, and curves are used to suggest temporal evolution or spatial transformations~\cite{davis2004informal}. Together, these practices enable people to represent sequences of motion and interaction without requiring full implementation, offering a lightweight yet powerful foundation for subsequent production.

Building on these practices, research has explored how to interpret users' authoring intent from sketches and translate that intent into final content.
For example, generative models can add and remove objects in a video based on a sketch drawn on a single video frame~\cite{Sketch3DVE2025Liu, liu2024generativevideopropagation}, or generate an entire video sequence from a storyboard sketch~\cite{liu2025sketchvideo,zhong2025sketch2anim}.
Beyond generation, sketches have also been explored for incremental authoring~\cite{landay1995interactive,davis2008ksketch,kazi2016motion}, where users construct animations step by step and each stroke is recognized~\cite{wobbrock2007gestures} or matched to a predefined command~\cite{davis2008ksketch}, triggering corresponding system actions (e.g., appearing, translating, or morphing) that gradually accumulate into a complete animated result.
However, these approaches often constrain the expressive potential of sketches, limiting them either to static layout representations or to predefined visual forms that encode specific animation commands. As a result, sketches shift from being an approach for open-ended creativity to functioning as rigid control tokens, missing the opportunity to harness their flexibility and spontaneity for expressing free-form dynamic intent, such as motion and interaction~\cite{tseng2011howuncertainty}. 

We aim to investigate how free-form sketches can effectively capture dynamic intent and be leveraged as a foundation for generating dynamic content.
Fortunately, recent advances in vision–language models (VLMs) have introduced powerful visual understanding capabilities, including the ability to interpret abstract and informal sketches. Such capabilities have been demonstrated in recent works such as CodeShaping~\cite{yan2025codeshaping}, though primarily in the domain of code generation.
We select motion graphics as our starting point, as they offer accessible yet expressive animations with clear motion dynamics, making them a practical setting for investigating the alignment between sketches and generated animations. 
\revisel{Specifically, we focus on explainer-style motion graphics with the roles of teaching aid and visual discourse~\cite{chevalier2016animation}, which are widely used across diverse domains and age groups~\cite{kazi2014kitty}, prioritizing the clarity of motion over visual realism to effectively communicate concepts~\cite{jahaolou2022katika}.}

Building on this foundation, we conduct a three-stage study to explore how sketch-based interactions can drive dynamic content creation.
In the first stage, to study sketch expressiveness and VLM interpretation, we provide a unified web interface where users can draw free-form sketch storyboards to generate animated videos through code-based rendering. The results highlight the strong potential of free-form sketches to convey animation intent with minimal, intuitive input \revise{(see Fig.~\ref{fig:sketch_example} for a collection of such examples)}, though their inherent ambiguity often leaves user intention underspecified. To address this, the second stage introduces a clarification cue that adaptively matches sketch ambiguity level, offering lightweight interface interactions for users to intervene in the VLM’s interpretation. Findings suggest that clarification turns sketch ambiguity into a resource for intent articulation, but some intentions only become concrete when users view the generated output and refine it accordingly. Therefore, in the last stage, we develop an intuitive visual editing approach that allows users to express their intent on top of the generated result directly. The refinement mechanism closes the gap between system output and user intent, allowing precise control with low effort.


In general, rather than pursuing high-quality video results, our contribution lies in reactivating the expressive freedom of sketches, which are casual and intuitive drawings that anyone can produce to freely convey dynamic intent without any constraints. Here are our contributions:

\begin{itemize}[left=0pt]
    \item \revisel{We propose a sketch-based dynamic intent expression paradigm where sketches are not predefined forms or fixed commands, but as open-ended prompts, in which sketch ambiguity, VLM interpretation, and user intervention collaboratively shape dynamic content.}
    \item \revisel{We implement \tool, a proof-of-concept instantiation of this paradigm, using free-form sketch storyboards to generate explainer-style motion graphics, supported by an adaptive clarification interface and multimodal input for iterative refinement.}
    \item \revisel{Through a three-stage user study, we validate the effectiveness of this paradigm. We demonstrate its capacity to support diverse free-form sketches, enabling users to articulate precise dynamic intent with minimal interaction.}
\end{itemize}

\section{Related Work}

\subsection{Understand Authoring Intent in Sketch}

Sketching has long been valued for its ability to convey authoring intent through just a few strokes, making it a widely adopted practice in early-stage ideation across fields such as architecture~\cite{Yıldızoğlu2024sketching,designmemo2025Li}, art, animation, and user interface design~\cite{landay1995interactive}. Numerous HCI researchers have explored how sketches can produce meaningful results for rapid prototyping, emphasizing their immediacy and fluidity as a medium~\cite{chen2023autocomputational}.

\begin{figure*}[ht!]
  \centering 
  \includegraphics[width=1\linewidth]{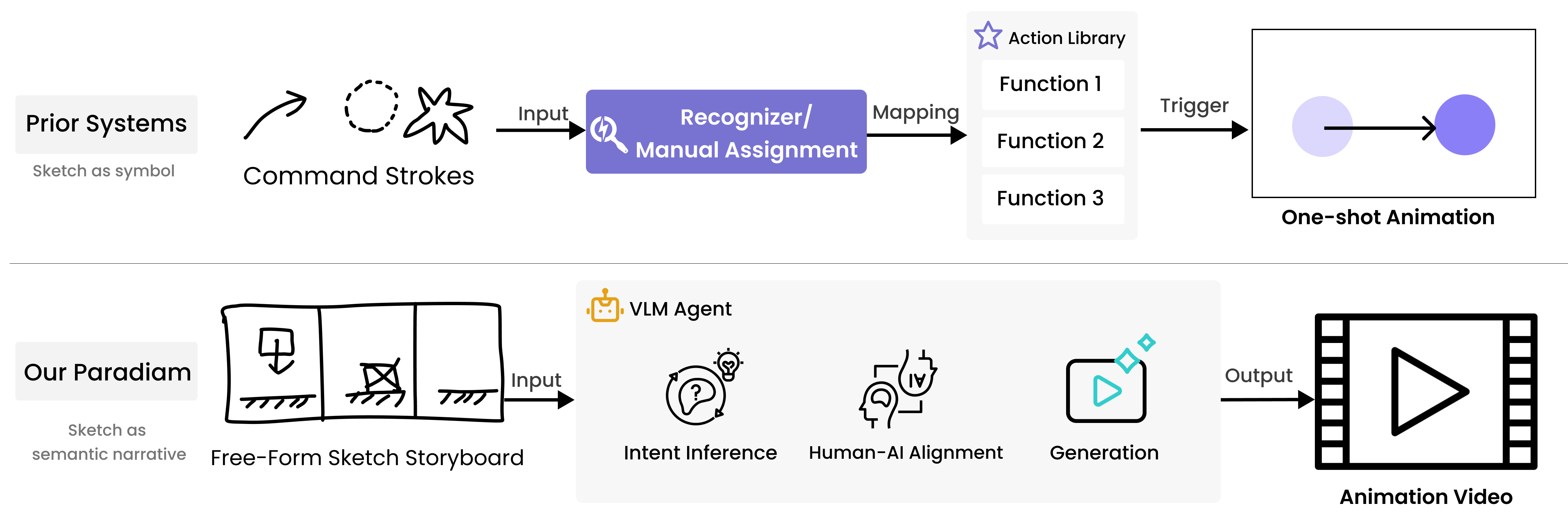}
  \caption{%
    \revise{Prior systems~\cite{davis2008ksketch,kazi2016motion,kazi2014draco,kazi2014kitty} treat sketches as symbols to be mapped to animation commands (Top). In contrast, we treat sketches as semantic expressions of motion intent (Bottom), leveraging VLM-based commonsense reasoning and human-AI clarification to interpret why and how motion happens.}
  }
  \label{fig:RW}
  \Description{}
\end{figure*}

A major part of this research treats sketches or single strokes as predefined symbols or commands~\cite{wobbrock2007gestures,rubine1991specifying, yuan2024generating}. With simple stroke recognition methods~\cite{tracy2008freesketch} like the \$1 recognizer~\cite{wobbrock2007gestures}, the system could classify sketches and trigger specific operations. A seminal example is SILK~\cite{landay1995interactive}, which allowed interface designers to hand-draw UI widgets and manipulate them with drawn strokes. Lineogrammer~\cite{zeleznik2008lineogrammer} extended this idea by beautifying sketches into clean diagrams and introducing more flexible action sketches, such as joining lines or lasso selection, enabling intuitive manipulation without tool switching. Using a similar technical pipeline, K-Sketch~\cite{davis2008ksketch} introduces an interface that enables novice users to create 2D animations through informal sketches. These interfaces provided immediate feedback by recognizing drawn shapes as components and commands, preserving the speed and informality of sketching. 
\revisel{However, as shown in Fig~\ref{fig:RW}, these systems interpret sketches through rigid one-to-one mappings, translating recognized symbols (e.g., arrows, lassos) directly into predefined graphical transformations (e.g., translation, rotation) or specific animation effects (e.g., morphing, fading). To support such mappings, designers needed to predefine gesture categories and their corresponding actions, meaning that users had to learn and conform to the system’s sketch–action vocabulary. As a result, while interaction felt informal and fluid, the semantic space of motion intent remains constrained to what had been explicitly encoded, preventing sketches from expressing abstract, causal, or contextual forms of motion intent.}

Therefore, some research has moved beyond hand-coded sketch mapping, learning to infer higher-level intent behind a sketch~\cite{kang2017structured}. Instead of asking ``what symbol was drawn?'', these systems learn to ask ``what does the user mean by this drawing?'' In user interface design, Swire~\cite{huang2019swire} exemplifies this shift by training a neural network on thousands of drawn UIs: it can interpret a free-form sketch of a UI layout in terms of its semantic similarity to existing interface examples. Samuelsson et al.~\cite{samuelsson2020eliciting} conducted an elicitation study with software engineers to understand how sketches might express code editing commands. Building on such insights, Code Shaping~\cite{yan2025codeshaping} introduced an interactive environment where developers draw directly over their source code to invoke edits. Behind this is the ability of current vision–language models: rather than simple classification, such models can understand quite abstract sketches, interpret them, and generate new content.

However, VLMs cannot always understand what users want to express in free-form drawings. Beyond the limitations of current algorithms for sketch understanding, an important reason is that sketches inherently carry ambiguity~\cite{tseng2011howuncertainty}. Some prior work has addressed this by imposing additional constraints to guide the AI’s interpretation. For example, Code Shaping ultimately introduced \q{command brushes,} a restricted set of sketch shapes each mapped to a specific edit operation, in order to improve recognition reliability. While this approach improved accuracy, it effectively reintroduced a symbolic shorthand, thereby limiting the expressive freedom of sketching. In contrast, we view ambiguity not as a flaw to be eliminated but as a feature that can drive creative workflows~\cite{gaver2003Ambiguity}. Our approach further explores engaging users in co-interpretation, where the system proposes an interpretation and the user iteratively refines or corrects it. This paradigm is especially well-suited to animation and dynamic media authoring, where sketches are inherently provisional and iterative refinement aligns with how creators naturally explore temporal ideas~\cite{davis2004informal}.

\subsection{Sketch as a Generation Prompt}
Sketches have been widely explored as prompts for generative algorithms due to their intuitive and expressive nature. Since sketches provide a quick and intuitive way to communicate spatial structure and layout, they have been widely used to guide the generation of static content such as images~\cite{chen2020deepFaceDrawing,zhang2023controlnet,xiao2021sketchhairsalon,lin2025sketchflex}, 3D models~\cite{lun2017shape,luo2024sketchmetaface}, websites~\cite{arefin2023pic2code,sonje2022draw2code}. 
Building on these applications, recent work~\cite{jin2024sketchguidedmotiondiffusionstylized,kling,liu2024generativevideopropagation,zhong2025sketch2anim,liu2025sketchvideo} has begun to explore the use of sketches in dynamic content generation, where conveying temporal and semantic change becomes essential. 
In pixel-based video generation, sketches can be drawn directly onto images to indicate motion trajectories or directional flow, serving as intuitive cues for generating animated sequences~\cite{kling}. \revise{Alternatively, sketches are used as masks or editing guides, allowing users to specify which regions of a video should change and how those transformations evolve over time time~\cite{liu2024generativevideopropagation,tilekbay2024expressedit,yang2024direct}.}

While sketch cues overlaid on images are effective for indicating short-term motion or localized edits, recent works~\cite{landay1996sketching,guay2015sketching,davis2006asketch,Garcia2019Spatial} have begun exploring sketch storyboards to support longer-range structure and narrative purely through sketch.
Traditionally used in early-stage film and animation ideation, sketch storyboards offer a sequence-based representation of motion and narrative intent, making them a natural fit for guiding generative algorithms~\cite{hart2013art}.
Different systems adopt sketch storyboards in domain-specific ways. SketchVideo\cite{liu2025sketchvideo} treats sketches as static illustrations of each keyframe’s visual structure, using carefully drawn outlines to guide photorealistic image generation. Sketch2Anim\cite{zhong2025sketch2anim}, on the other hand, relies on a predefined sketch format, typically a stick figure combined with a trajectory, to represent skeletal movement across frames. In both cases, the sketch input is tightly constrained and tailored to the specific task.

While effective for domain-specific tasks, these constrained sketch representations fall short for motion graphics video creation, where compositions often feature multiple vector-based elements moving in abstract, stylized, or non-physical ways~\cite{krasner2013motion,brinkmann2008art}.
Such sketch formats impose rigid structural or pose constraints, limiting users’ ability to articulate complex and unconventional animation ideas~\cite{yan2025codeshaping}.
To overcome these limitations, we adopt free-form sketch storyboards that support open-ended drawing without constraints, enabling richer expression of spatial, temporal, and stylistic intent.
This flexibility allows sketches to function not as fixed blueprints but as adaptable cues that guide generation. Building on this foundation, our system further supports an interactive, iterative authoring process, allowing users to refine their animations over multiple rounds rather than relying on one-shot generation.

\subsection{Vector Animation Authoring for Casual Creators}
Creating motion graphics or vector animations traditionally requires substantial expertise and professional tools such as Adobe After Effects~\cite{ae}. HCI research has long sought to lower this barrier by developing more intuitive authoring paradigms. Early sketch‐driven systems such as Draco~\cite{kazi2014draco} and Kitty~\cite{kazi2014kitty} let users draw scenes and motion cues directly on the canvas, inferring causal animations without requiring keyframe editing, while SketchStory~\cite{lee2013sketchstory} applied similar ideas to animated data storytelling. These works emphasize continuous, pen‐on‐canvas interaction over traditional timelines. Building on this, more recent end‐to‐end systems scaffold the entire workflow for novices: Katika~\cite{jahaolou2022katika} guides users from storyboards or scripts to finished videos via libraries of pre‐designed animations, and DancingBoard~\cite{chen2025dancingboard} streamlines motion‐comic production with guided steps and automated animation suggestions.
\revise{Some works focus on more specialized domains of vector animation.}
In data visualization, tools such as Data Illustrator~\cite{liu2018dataillustrator} and Data Animator~\cite{thomson2021dataanimator} enable designers to create animated charts as if drawing static graphics, automatically interpolating between states. \revise{For text animation, works like The Kinetic Typography Engine~\cite{lee2002kinetic} and TextAlive~\cite{kato2015textAlive} demonstrated powerful systems where users could algorithmically define and control motion.  Piet~\cite{shi2024piet} further extends this direction by providing targeted interfaces that facilitate color authoring in motion graphics videos.} Most recently, AI‐driven authoring has emerged: Keyframer~\cite{tiffany2024keyframer} and AnyAni~\cite{qiu2025anyaniinteractivegenerativeai} use large language models to translate natural‐language prompts into editable vector animations, supporting iterative refinement through text or direct manipulation. 

Building on this trajectory, researchers have begun to explore how vision–language models (VLMs) can support motion graphics generation. MoVer~\cite{ma2025mover} introduces a motion verification DSL that encodes spatio-temporal properties of animations in first-order logic, demonstrating how formal reasoning can be applied to ensure correctness of motion behaviors. LogoMotion~\cite{liu2025logomotion} analyzes vector graphics (SVGs) to automatically suggest motion designs, showing how pre-existing assets can be transformed into animations.
These works highlight the potential of AI models to reason about both semantics and temporal logic in vector-based media.
However, a gap remains. Existing systems have largely focused on short, self-contained examples with relatively few animated elements, often suited for toy demos or tightly scoped use cases. In contrast, many real motion graphics videos—even simple explainers—already involve multiple objects evolving over time, which introduces additional complexity for authoring and interpretation. Supporting such scenarios requires more direct and flexible visual guidance, enabling creators to express dynamic intent without relying solely on pre-defined templates or textual prompts. In this paper, we explore free-form sketches as a visual guidance mechanism. Sketches offer an intuitive way to externalize spatial–temporal ideas, while VLMs can interpret these abstractions into vector animations. This combination bridges informal creative expression with structured animation output, \revise{extending bi-directional editing~\cite{hempel2019sketch-n-sketch} to non-programmers.}

\begin{figure*}[ht!]
  \centering 
  \includegraphics[width=1\linewidth]{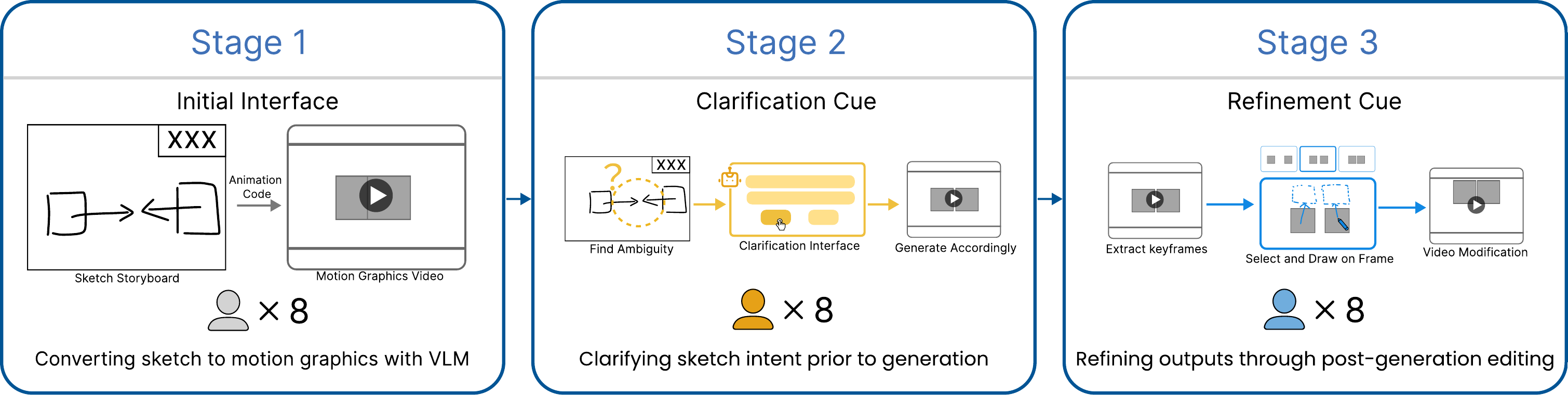}
  \caption{%
    Study illustration showing the three stages of our investigation: (1) sketch input and direct generation, (2) clarification through disambiguation cues, and (3) iterative refinement with contextual editing. Each stage involved eight new participants in the study.
  }
 
  \label{fig:overview}
  \Description{}
\end{figure*}
\section{\tool}

Free-form sketch storyboards are often used for quick idea exploration in dynamic content creation, and they also hold promise as input for generation. Yet it remains unclear how such informal sketches can reliably convey a creator’s intent across the authoring process. To explore this, we developed a proof-of-concept system, \tool, which examines how free-form sketches can function as a medium for expressing animation ideas. 
With this system, users sketch simple storyboards, AI interprets the drawings, translates them into vector-based animation code, and renders them into video.
To situate the study in a prototyping context that mirrors early creative practice, we did not require participants to have advanced drawing skills, encouraging them to use simple and intuitive sketches within just a few minutes as a way of communicating their animation intentions.

We design the system iteratively through the following three stages (see Fig~\ref{fig:overview}):
\subsubsection*{\textbf{Stage 1: Understanding sketch and interpretation.} In this stage, we implemented a unified interface that understood sketch storyboards from users and transformed them into motion graphic videos (Sec.~\ref{sec:stage1_initial}). We explored how participants naturally expressed animation intent through free-form sketches and how these sketches were interpreted when directly converted into videos. The study findings revealed that participants tended to use highly diverse and abstract sketches for quick prototyping. However, even though the VLM demonstrated impressive capacity to interpret many inputs, it still often failed when confronted with ambiguous sketches.}
    
\subsubsection*{\textbf{Stage 2: Investigating ambiguity in sketch interpretation.} Through empirical use in the first stage, we identified recurring cases where user sketches were too abstract or ambiguous for the system to interpret reliably (e.g., an arrow could indicate either translation or rotation). This led us to investigate a lightweight way for enabling users to intervene in the machine's interpretation process. In this stage, we proposed a clarification cue mechanism (see Sec.~\ref{sec:clarification}), which separated ambiguity into four levels and provided adaptive interfaces for user input. Our study showed that participants generally considered clarification requests reasonable and helpful, noting that such prompts not only reduced misinterpretation but also assisted them in further shaping their intent.}
    
\subsubsection*{\textbf{Stage 3: Exploring more intuitive iterative editing.} Not all ambiguities could be resolved in the early stage, as some sketches were too underspecified to be faithfully interpreted until a video was generated to provide additional context. In this stage, we investigated how users could refine the output when it did not meet their expectations, introducing a frame-based interaction method (see Sec.~\ref{sec:refine}) that combined keyframe extraction and annotation to offer clearer guidance toward the desired result. Our study showed that the refinement strategy was efficient and intuitive, allowing participants to annotate video frames to express intent with minimal effort.}


\begin{figure*}[ht!]
  \centering 
  \includegraphics[width=1\linewidth]{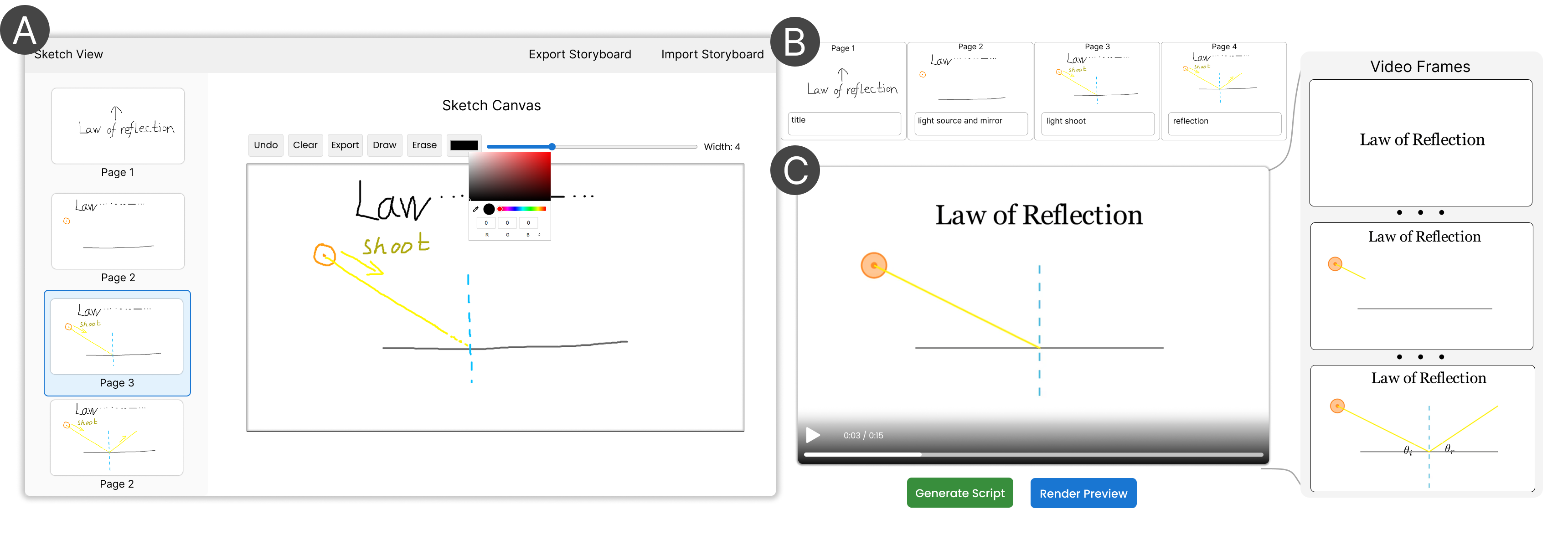}
  \caption{%
    Interface design for the first Stage. (A) The user sketches a frame-level storyboard to explain \revise{the law of reflection, drawing incident and reflected light rays on a mirror surface}. (B) adds brief notes at the bottom to specify the cues, and (C) generates a preview video from the storyboard.
  }
 
  \label{fig:exploration1}
  \Description{}
\end{figure*}

\section{Stage 1: Initial Sketch to Animation Authoring}
\revise{We began with a user study to explore how participants sketched animation ideas and how these sketches were translated into preliminary video outputs. Rather than optimizing for output quality, this baseline stage focused on observing sketching behaviors and interpretive patterns, laying the foundation for identifying issues and guiding subsequent investigations. We chose this approach over an artifact-based analysis of existing online storyboards~\cite{davis2008ksketch}, as the uncontrolled nature of such materials (e.g., varying time and quality) would have limited the validity and comparability of our observations.}

\subsection{Initial Implementation}
\label{sec:stage1_initial}

To begin with, we designed a unified web interface where users can directly explore how their free-form sketches translate into motion graphics. 
The interface consisted of three connected views (Fig.\ref{fig:exploration1}).
In the sketch view (Fig.~\ref{fig:exploration1}-A), users drew on a central canvas using freehand strokes. The toolbar at the top provided lightweight editing functions such as undo, clear, pen, eraser, color, and pen size. Sketches could be exported or imported in multiple formats (PNG, SVG, or JSON data), and entire storyboards could be saved or reloaded. A sidebar displayed all sketches as page thumbnails, allowing users to navigate between frames or remove sketches.
Once sketches were created, they were organized in the storyboard view (Fig.~\ref{fig:exploration1}-B).  The interface displayed all sketches in sequence, giving users an overview of how their drawings formed a storyboard. Each sketch could optionally be annotated with short text notes, but the interface emphasized sketches as the primary mode of input.
When the storyboard was complete, the user entered the AI video view (Fig.~\ref{fig:exploration1}-C). Here, a single click on [Generate Script] sent the storyboard to the model, which returned executable code. By pressing [Render Preview], the code was compiled into video and displayed in the built-in player, allowing the user to immediately check the generated result.

Behind the interface, we carefully engineered the sketch-to-video translation process to support this interaction. The first step was ensuring that the VLM could interpret sketches and generate corresponding animation code. To this end, we designed prompts with paired sketch descriptions, code examples, and formatting rules, asking the model to output executable scripts. A central challenge was how to represent the storyboard so the model could reason across sketches effectively. We tested two input strategies: (1) feeding each sketch individually with its script, and (2) compositing all sketches and text scripts into a single storyboard input. Results indicated that the second approach produced more coherent and temporally consistent animations, likely because the model could better reason across sequential sketches. This strategy was therefore adopted in subsequent development.
The second step was to render the generated code into video. We instructed the VLM to output executable Manim~\cite{manim} Python code, which was then compiled into animations. Manim, a declarative animation library widely used for educational explainer videos, provided the rendering backend. This code-first design not only enabled fine-grained control at the script level (e.g., timing, trajectories, styles) but also ensured that animations were built from scalable vector-based primitives, making the video extensible and well-suited for future iterative editing.

\subsection{Study Design}
\subsubsection*{\textbf{Participants}}
Consistent with previous research~\cite{yan2025codeshaping}, we recruited eight participants (P1–P8; five male, three female) from a local university, with an average age of 26.2 years (SD = 4.2). All participants were right-handed and had prior exposure to motion graphics as viewers. Five participants reported previous experience in creating motion graphics using tools such as After Effects~\cite{ae} or CapCut~\cite{capcut}, while the remaining three had only basic video editing experience, primarily limited to simple cutting and sequencing.
We intentionally recruited participants with varying levels of expertise to capture perspectives from both experienced and novice creators, while ensuring that all had sufficient familiarity with motion graphics as consumers to provide informed feedback on the generated results.

\subsubsection*{\textbf{Tasks}}
Since our system was designed to support storyboard creation from scratch, we did not impose a narrowly constrained task or theme. Instead, we asked participants to imagine creating a short explainer video for a concept of their choice. To keep the task manageable while still showcasing the workflow, we provided three guidelines: (1) the video should illustrate a concept that can be reasonably explained through basic shapes, (2) participants should avoid overly complex shapes or animation techniques, and (3) the storyboard should be concise, consisting of around four sketches. \revisel{Specifically, we limited concepts to logical explanations to evaluate functional accuracy; We restricted visuals to basic shapes to abstract away illustrative details, allowing participants to focus their cognitive effort on articulating dynamic logic (e.g., causality, flow) rather than static aesthetics; and we focused on fundamental animation primitives (e.g., displacement, deformation) rather than high-dimensional tasks like character animation, as their complexity makes it difficult to verify the specific animation intent.} We conducted the experiment on a desktop interface with mouse-based interaction, as our focus was not on the precision of sketching.

We restricted visuals to basic shapes to abstract away illustrative details, allowing participants to focus their cognitive effort on articulating dynamic logic (e.g., causality, flow) rather than static aesthetics
\begin{figure*}[h!]
  \centering 
  \includegraphics[width=1\linewidth]{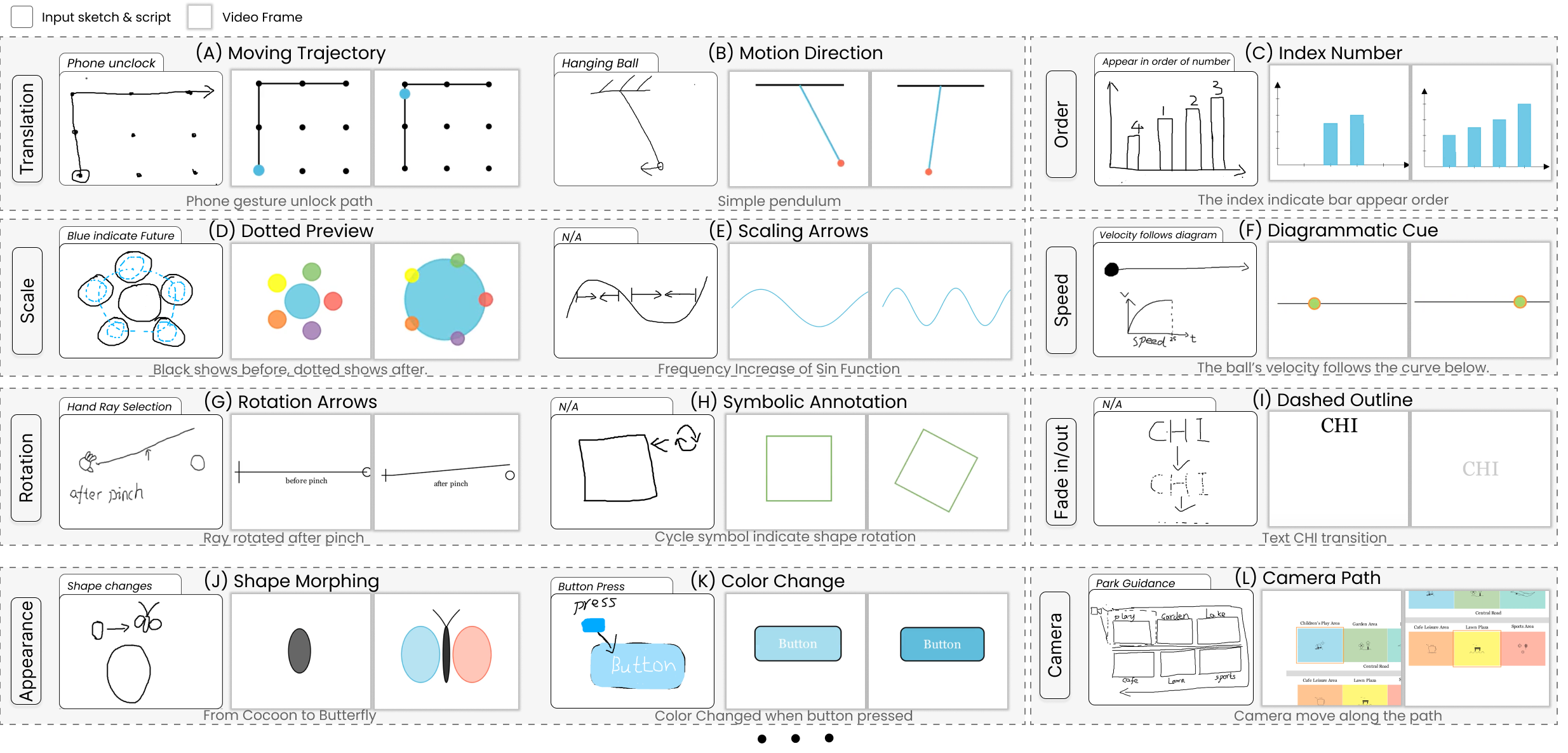}
  \caption{%
    Selected sketches with scripts and animation excerpts from participants’ storyboards in Stage 1. Above the sketch is the script(N/A means no script); the gray text below describes the animation. These examples illustrate common ways sketches were used to convey animation intentions (e.g., translation, scaling, rotation, appearance). The categories shown here are not meant as a strict taxonomy but as illustrative examples; in practice, participants’ free-form drawings were far more varied and often mixed multiple notations within a single sketch. Some animations shown are refined versions to better reflect participants’ intended outcomes.
  }
 
  \label{fig:sketch_example}
  \Description{}
\end{figure*}
\subsubsection*{\textbf{Procedure}}
Participants began the experiment by completing a survey to gather information on their prior experience with video creation and sketching (Appendix.~\ref{sec:appendix-survey}). Next, they were given a brief overview of the project and shown the core features of the interface. Some example cases were presented to illustrate the expected input and outcome. We specifically presented examples of free-form sketches and encouraged participants to use more flexible sketching methods to express their ideas, with less emphasis on scripting. Following the introduction, participants had a 10-minute exploration phase, where they could experiment with the sketching tools, test the generation of animation previews, and become familiar with the interface's capabilities. 
Participants then moved on to the creation phase, where each participant was tasked with creating three separate videos, each representing different content. This meant that they needed to create three distinct sketch storyboards and review three generated videos, completing the task in approximately 30 minutes.
Finally, the session concluded with a semi-structured interview and a questionnaire, where participants provided feedback on their experience in translating ideas into sketches, as well as their thoughts on the generated videos.
\revise{The similarity between participants’ intended animations and the generated outputs was analyzed using both qualitative feedback and the Alignment metric from the questionnaire.}

\subsection{Results}

Across all sessions, we collected 24 attempts. Not all were successful: five outputs were judged by participants as failures, where the generated video significantly diverged from their intended idea. As a first-stage probe, such outcomes are acceptable and highlight the current gap between sketch expressiveness and model interpretation.
Participants appreciated that they were allowed to attempt animations across a wide variety of domains, rather than being constrained to a specific area.
The topics participants chose to explore were diverse, covering but not limited to VR interaction, data visualization, phone interaction, traffic rules, celestial motion, algorithm pipelines, mathematical concepts, physical rules, and logo animation. The unsuccessful cases provide further insight into the current limitations of sketch-based intent expression. These included a block colliding back and forth with a wall (P2), a maze-walking path (P2), a rainfall animation (P5), a bouncing ball (P7), and a grid-warping effect (P8). Based on different animation intents and sketch types, we selectively demonstrate some results from the user in Fig~\ref{fig:sketch_example}.

\subsubsection*{\textbf{Expanding the Role of Sketches}}
In our study, sketches served not only as a way to specify layout or static elements, but also as a medium for expressing animation intent. When attempting to represent dynamics, participants resorted to a wide variety of abstract conventions such as arrows (Fig.~\ref{fig:sketch_example}-B), \revise{onion skinning} for future states (Fig.~\ref{fig:sketch_example}-D), or index numbers for sequencing (Fig.~\ref{fig:sketch_example}-C). However, these conventions were highly diverse and often idiosyncratic, shaped by individual habits and prior experiences. For instance, P3 emphasized that \q{everyone has their own way of drawing arrows, mine is from engineering class}, while P6 noted that \q{sometimes I just use circles, sometimes a line, it depends on what comes to mind first}.  
This diversity meant that the same animation could be represented in multiple ways, and similar sketches could point to different intentions. As P4 reflected, \q{the arrow could mean movement, or just attention, it depends on the context}. Indeed, 7 out of 8 participants explicitly remarked that they have their own strategies to depict movement, often \q{no regulation} (P2) or \q{quite random} (P5). This means that free-form sketching opened up a broad expressive space, but one that was highly individualized and semantically unstable, requiring interpretation and negotiation when used to communicate temporal behaviors. 

\subsubsection*{\textbf{Sketch Interpretation Beyond Expectations.}}
Participants often experimented with sketches that were highly abstract or loosely defined, yet many of these were still interpreted meaningfully by the model. Several participants created inventive sketch forms to convey intent—for example, using \revise{gesture line} or symbolic marks to indicate motion or transformation. Despite this, 6 out of 8 participants admitted that, at the beginning, they doubted whether such free-form drawings could be understood by the machine. P8 further reflected that some sketches were inherently ambiguous, \q{different people could read it differently.} While some outputs indeed diverged from their expectations, participants often adapted by redrawing their sketches or supplementing them with short text scripts to clarify intent. Interestingly, in 7 cases, participants reported being positively surprised when the system successfully captured the essence of their sketches in ways they had not anticipated. 
\revise{Although the user ratings (Fig.~\ref{fig:rating}) indicate that most participants felt the animations were not a perfect match to their intended outcome, they still perceived the results as capturing their intent to some extent and described them as \q{unexpected but meaningful}.}

\subsubsection*{\textbf{Semantics over geometry.}}
A recurring theme was that the VLM often prioritized semantic intent rather than literal geometric fidelity. Instead of replicating sketches exactly, the model tended to generate canonical or cleaned-up versions of the intended motion. For instance, when P6 drew a wobbly sine curve, the output appeared as a smooth sinusoid. P6 described this as \q{beautified}, appreciating that the messy strokes became clearer in the final animation. Similarly, P9 noted that their \q{rough arrows} were rendered as consistent trajectories, which they found \q{nicer than I expected.}
However, this abstraction also created moments of mismatch. P4, who attempted to specify a very precise trajectory, complained that the output \q{ignored the exact angle I drew}. Likewise, P11 commented that \q{it feels like it understood the idea but not the detail}. In total, 7 of 8 participants remarked that the system often generalized their sketches to capture meaning rather than form, with reactions ranging from surprise and delight to mild frustration. Several (e.g., P3, P5) responded by redrawing shapes or adding text notes to enforce precision.

\subsubsection*{\textbf{Balancing detail and abstraction in sketching.}}

Participants varied in how much detail they invested in their sketch storyboards. At one extreme, some treated the canvas almost like a frame-by-frame storyboard, drawing each scene carefully to make their intent explicit. While this approach produced clearer outputs, participants also described it as \q{tedious} (P3) and \q{too much effort for a short video} (P7). In contrast, many preferred using abstract marks that conveyed animation intent more quickly. As P6 explained, this was \q{the fastest way to show the idea,} even though it sometimes made the system’s interpretation less predictable. P8 similarly noted that abstract cues \q{save time but leave ambiguity.} This trade-off reflects a central tension: detailed sketches improve clarity but increase workload, while abstract sketches reduce effort but demand stronger interpretive support from the model. Reflecting this tension, P5 emphasized that although they were willing to rely on abstract sketches, they expected the system to \q{just understand the sketch} without needing additional textual specification.

\begin{figure*}[ht!]
  \centering 
  \includegraphics[width=1\linewidth]{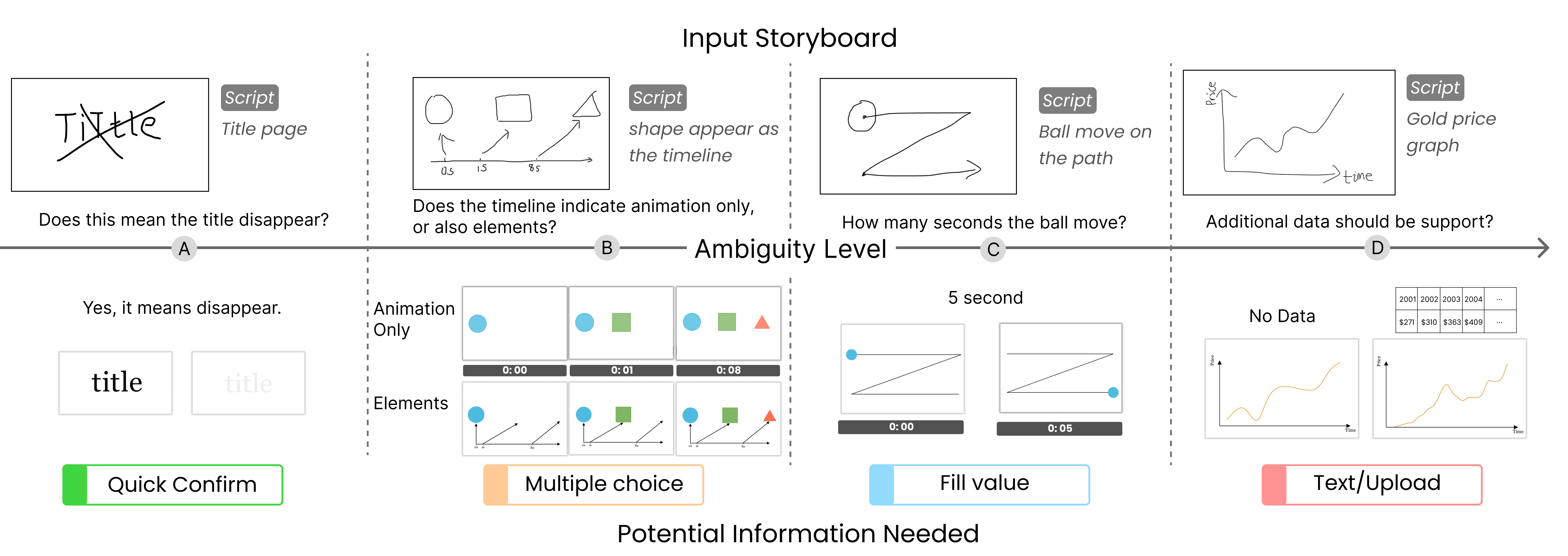}
  \caption{%
   Illustration of ambiguity clarification. From left to right, the examples show sketch ambiguities ranging from low to high. Top: different types of ambiguities in storyboards, Bottom: the corresponding types of information needed to resolve them, mapped to our clarification strategies.
  }
 
  \label{fig:ambiguity}
  \Description{}
\end{figure*}

\section{Stage 2: Clarification Guide Interpretation}

Findings from the first stage revealed two key challenges in sketch-based authoring. The first concerned the gap between a user’s intended animation concept and the sketch they produced. Participants often relied on highly abstract marks inspired by everyday experience or improvised on the spot; while efficient for the drawer, such sketches were difficult for others to interpret and often left their meaning ambiguous. The second challenge arose between sketch input and machine interpretation, where the limitations of current AI reasoning led to outputs diverging from participants’ expectations.
\revise{For example, in Fig.~\ref{fig:ambiguity}-B, the VLM might randomly decide whether to include the annotated timeline in the final animation, resulting in an output that does not match the user's expectations. Similarly, in the case of Fig.~\ref{fig:ambiguity}-D, the VLM might generate an arbitrary curve, whereas the user may have intended to create a visualization based on real data.}
To address these issues, the second stage aimed to involve users more directly in the interpretation process, allowing them to understand and influence how sketches were translated into animations. Crucially, this was designed without burdening users to provide lengthy textual descriptions. Instead, we developed lightweight mechanisms that let users guide and refine the system’s understanding interactively.

\subsection{Clarification Cue}
\label{sec:clarification}
Prior work has highlighted that sketches are inherently ambiguous~\cite{suwa1997what,barbara2009tools} and that such ambiguity can be a productive resource in design interpretation~\cite{gaver2003Ambiguity}. Our observations in the first stage confirmed this tendency in the context of animation authoring: participants often relied on abstract arrows, shorthand marks, or improvised symbols to convey motion (Fig.~\ref{fig:ambiguity}), which were efficient for them but difficult for others, or the VLM, to interpret. Even when the intended action was clear, essential details such as duration, scale, or speed were frequently left unspecified. These findings illustrate that ambiguity in sketching arises not only from multiple possible interpretations but also from underspecified parameters and context-dependent symbols, all of which can cause model outputs to diverge from user expectations.

Rather than treating ambiguity as an error to be eliminated, we approached it as a variable condition to be managed. Our design philosophy was to preserve the freedom of freehand sketching while offering just-in-time interventions to clarify intent when needed. 
\revise{This philosophy echoes early ideas of free-form sketch beautification~\cite{igarashi1997interactive},  supporting multiple candidates for ambiguity and requesting the user's interaction}.
To this end, we adopted a layered prompting strategy that adaptively maps the degree of ambiguity to different levels of system intervention. This draws on the principle of progressive disclosure: users only encounter as much clarification effort as the ambiguity of their sketches demands, thereby minimizing disruption while maintaining interpretability.
Concretely, based on previous research~\cite{davis2008ksketch,avola2010classifying} about sketch ambiguity and empirical results from the first stage, we developed four complementary mechanisms that align with increasing levels of sketch ambiguity, ranging from low to very high, each offering progressively richer forms of clarification, as shown in Fig.~\ref{fig:ambiguity}.
\begin{itemize}[left=0pt]
    \item \textit{Quick confirm}. For low-uncertainty cases such as uncertain strokes or vague geometry~\cite{barbara2009tools}, the system generates a primary guess and asks the user for yes/no confirmation (e.g., ``Should this line be used as a motion path?''). 
    \item \textit{Multiple choice}. When a sketch could plausibly mean several different things~\cite{gaver2003Ambiguity}, the system presents alternative previews and asks the user to pick the intended one (e.g., a curved arrow could be shown as either ``rotation'' or ``decorative arrow'').
    \item \textit{Fill value}. For actions that are recognized but lack specific parameters, the interface asks for numeric or scalar values. While many aspects of the video could in principle be parameterized, we prompt only for those with a significant impact on the outcome. For example~\cite{davis2008ksketch}, the system may ask how long a ball should take to traverse a path, while leaving minor details such as fade-in duration to sensible defaults.
    \item \textit{Text or upload resources}. In cases where sketches are highly abstract or symbolic~\cite{gaver2003Ambiguity,tseng2011howuncertainty}, the system allows users to provide free-form clarification, either as a short text note or by uploading an external reference asset.
\end{itemize}


\begin{figure*}[h!]
  \centering 
  \includegraphics[width=1\linewidth]{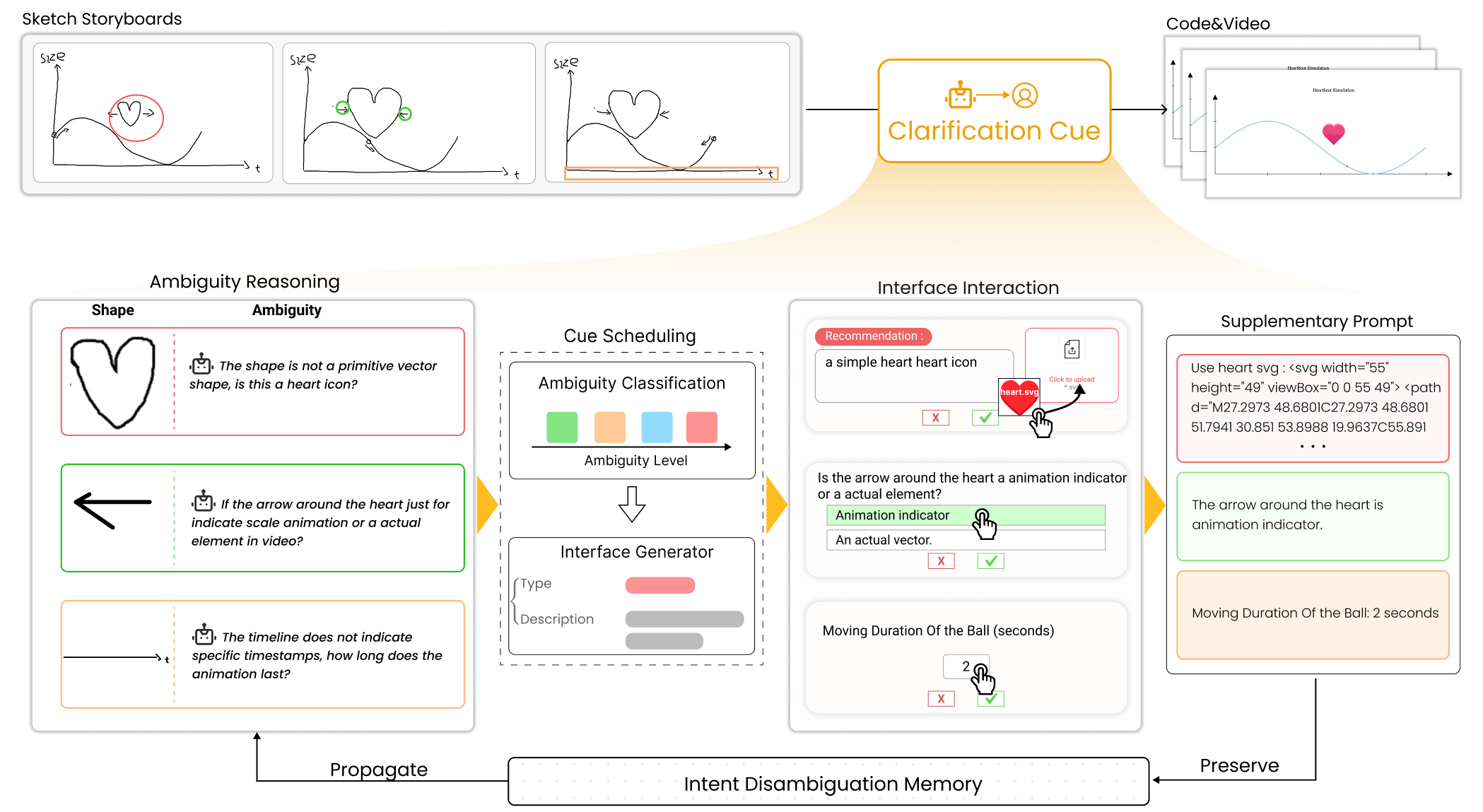}
  \caption{%
    Example workflow of the clarification cue mechanism. The system detects ambiguities in sketches, classifies them by level, prepares them as JSON for the interface generator, and prompts users with lightweight clarifications (e.g., confirmation, choice, value input, or asset upload). User responses are used as a supplementary prompt for regeneration and stored in an intent disambiguation memory to avoid repeated queries in future generations.
  }
 
  \label{fig:clarification_cue}
  \Description{}
\end{figure*}

After completing a sketch storyboard, users proceed to generation in the AI video view. When they press [Generate Script], if the VLM confidently interprets the sketches, the interface produces the animation code and renders a preview video directly. When ambiguities are detected, however, the interface does not produce arbitrary results; instead, it pops up a clarification panel. For example, if a curved arrow could be read as either rotation or decoration, the interface presents two animated previews for the user to select the intended one. In another case, if a user’s symbol is too abstract, they can upload a custom SVG asset (e.g., a heart icon to represent an object appearing with a ``pop'' effect).
For implementation, as shown in Fig~\ref{fig:clarification_cue}, the clarification mechanism is supported through prompt engineering and contextual memory. The VLM is explicitly instructed to request clarifications only when necessary and only on aspects that critically affect the generated animation, avoiding excessive or redundant queries. When the user provides input, such as selecting between preview options or uploading an external SVG asset, these responses are added as a supplementary context for regeneration. In addition, the system maintains an \textit{intention disambiguation memory}, which records user-provided clarifications and automatically reuses them when the same sketch is edited and regenerated later. 

\subsection{Study Design}
We recruited eight new participants (P1-P8; six male, two female) from the same university, with an average age of 25.4 years (SD = 2.9). All participants had prior experience in creating videos, though none were experts. The study design followed the same task guidelines and procedures as in Stage 1, with the addition of the new clarification mechanism. 
When a cue requires external resources (e.g., SVG files), we allowed users to directly search online to retrieve the corresponding assets.
During the study, we manually recorded participants’ sketches, the generated results, and the clarification cues that were triggered throughout the process, and collected user feedback.

\subsection{Results}

We collected a total of 87 clarification cues across the 24 creation attempts.  In addition, participants were free to revise their sketches and re-generate videos within the same attempt; for analysis, we counted all cues triggered during that cycle under a single attempt. This occasionally inflates the count for sketches that went through several iterations, but more accurately reflects the total clarification effort needed to reach a satisfactory result. In two cases (T5 and T10), participants ultimately judged the final output as misaligned with their expectations, both involving physics-inspired animations such as bouncing or collision. Overall, participants expressed a positive attitude toward the clarification cues. Rather than perceiving them as interruptions, several described them as \q{helpful checks} (P3) or \q{a way to steer the system back on track} (P7). Even when additional questions extended the workflow, participants valued the cues as opportunities to make their intent explicit and avoid wasted generations.

\subsubsection*{\textbf{Overall Cue Usage}}
\begin{figure*}[ht!]
  \centering 
  \includegraphics[width=0.8\linewidth]{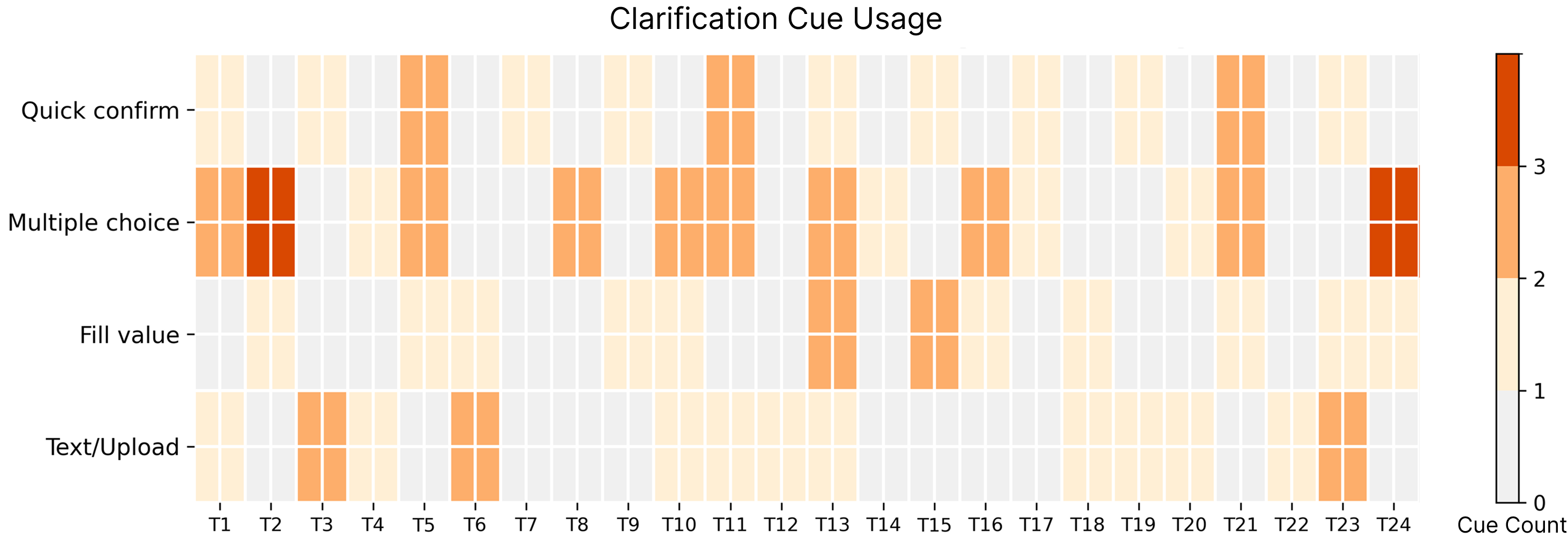}
  \caption{%
    Heatmap of clarification cue usage across 24 tasks (T1 to T24). The $y$-axis displays four types of clarification cues: Quick confirm, Multiple choice, Fill value, and Text/Upload. \revise{Each block in the grid represents the total Cue Count (frequency of use) for a specific cue type within a specific task. Color intensity corresponds to this count, with the scale shown in the legend on the right (darker orange indicates a higher count, up to 3).}
  }
  \label{fig:cue_usage}
  \Description{}
\end{figure*}

Figure~\ref{fig:cue_usage} summarizes the distribution of cue types per attempt in the form of a heatmap. The pattern shows that multiple choice cues were triggered most frequently, reflecting the prevalence of sketches with more than one plausible interpretation, which is the most common ambiguity. Quick confirm cues were relatively rare, as the model often resolved low‐ambiguity cases confidently without further questioning. Fill value cues also appeared infrequently because the system was designed to ask only for parameters with a major impact on the outcome, and that case is relatively uncommon. By contrast, text/upload cues occurred at a moderate level, indicating that participants frequently drew highly abstract or symbolic sketches (e.g., rough icons) that required extra resources or textual hints. In these cases, the prototype demonstrated its ability to reason semantically about the sketch and prompt the user for relevant supplementary information.

The heatmap also reveals strong variation across attempts. We did not include detailed per‐attempt breakdowns of cue types in the paper, as these would be too granular, but the heatmap variation and density of cue usage across the study are meaningful. Some creations, such as T5 and T13, required a dense sequence of cues (up to 5–6 within one attempt), while others (e.g., T7, T14) proceeded with almost no intervention. This disparity highlights how cue demand was closely tied to the degree of abstraction in the sketch. For instance, P3 commented that \q{I only needed help when I drew something unusual,} whereas P8 described the cue prompts as \q{a relief when I was not sure if the sketch was understandable.} 

\subsubsection*{\textbf{User Reactions to Cue}}
Participants generally responded positively to the clarification cues, with 7 out of 8 noting that they helped avoid misleading outputs and made them more aware of the ambiguity in their sketches. Quick confirm cues were described as \q{lightweight} and easy to dismiss, while multiple choice cues were seen as the most necessary since arrows and curves often admitted multiple interpretations. For example, in T10 a participant drew a looping arrow and appreciated being asked to choose between \q{rotation} and \q{decorative arrow.} Fill value cues divided opinions: some found specifying duration or speed helpful, while others felt it broke the free-flow nature of sketching. Text/Upload cues, though requiring more effort, were valued when rough sketches stood in for specific icons. In T23, a participant drew a rough star sketch, which triggered a clarification cue suggesting the upload of a pentagram icon. P8 remarked that they were \q{surprised} the system could interpret such an abstract mark as a symbolic request for a sparkle effect. 
\revise{Overall, all participants felt that the clarification cues made sense in resolving sketch ambiguities. According to the user rating (Fig.~\ref{fig:rating}), compared with stage 1, clarification cues generally align the user's intention with the final animations, and participants did not find these reasonable checks to be significantly disruptive to their creative process.}

\subsubsection*{\textbf{The Result after Cue}}
After clarification, 19 out of 24 attempts produced outputs that participants rated as closer to their intended animation, compared to the initial generation. In 7 attempts, users explicitly described the cue as the turning point that made the result usable (e.g., P6: \q{the timing became right only after I entered the value}). Fill value cues directly improved temporal qualities: in T13 and T21, entering a duration smoothed otherwise abrupt motions. Text/Upload cues often corrected semantic mismatches: in T23, replacing a hand-drawn star with an SVG asset transformed the effect into what P8 called \q{finally what I had in mind.} By contrast, Multiple choice cues mainly resolved directional intent; in T10, choosing \q{rotation} instead of \q{arrow symbol} shifted the system output to match the participant’s storyboard. Failures still occurred: 2 attempts involving physical dynamics remained unsatisfactory even after cues, suggesting limits in the model’s reasoning rather than the cue mechanism. Taken together, the logs show that cues did not simply add extra interaction but often directly determined whether a sketch led to a usable animation.

\subsubsection*{\textbf{Workflow Integration}}
Participants emphasized that the clarification cues were most valuable when integrated smoothly into the overall authoring workflow, rather than appearing as separate steps. In practice, cues were triggered inline during generation and resolved within a few clicks, which participants described as \q{not disruptive} (P3) and \q{just part of the flow} (P6). Several participants noted that without these cues, the system might \q{guess wrong} (P5) or return outputs that \q{look nothing like what I wanted} (P2). By contrast, lightweight confirmations or multiple-choice previews felt like natural checkpoints that improved trust in the system. Importantly, participants highlighted that the cues helped them understand how the model interpreted their sketch, giving them a sense of control. As P8 put it, \q{I can see what the AI thinks, and fix it right away.} Overall, rather than slowing down creation, clarification cues were seen as seamlessly woven into the workflow, balancing automation with user agency.

\section{Stage 3: Refinement with Visual Context}

In the second stage, we introduced clarification cues to help disambiguate sketches during the initial authoring process. However, once a video was generated, users who wanted to make changes could only revise their sketch storyboard and trigger a full re-generation. This workflow was inefficient and often unpredictable: re-generation could yield results that diverged widely from the original, even if the user only intended to adjust a small local detail. For example, participants described situations where they were satisfied with most of the video but wanted to tweak a single motion, only to receive a completely different animation after re-generation. To address this limitation, Stage 3 explores how users can visually express their intent for modification directly on the generated video,  refining the output step by step with iterative local editing.

\subsection{Refinement Cue}
\label{sec:refine}

\begin{figure*}[h!]
  \centering 
  \includegraphics[width=1\linewidth]{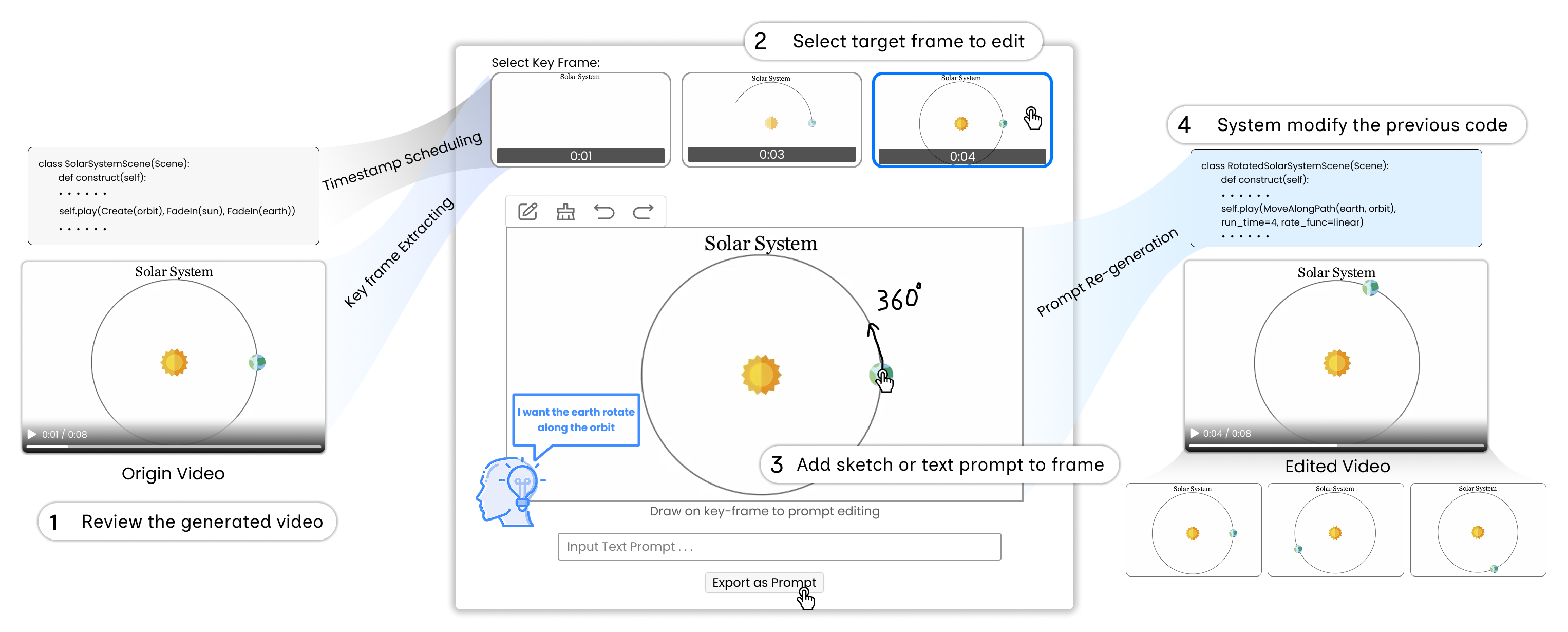}
  \caption{%
    Example of the refinement cue workflow. In this case, the user intends the Earth to move along its orbit. The process begins with (1) reviewing the initial video. Next, (2) the system automatically extracts a set of representative keyframes, from which the user selects the one most relevant to their intended edit. (3) On the chosen keyframe, the user adds an arrow to indicate the desired motion. Finally, (4) the system updates the underlying code accordingly and regenerates the video, producing an edited version where the Earth rotates along the orbit.
  }
 
  \label{fig:refinement_cue}
  \Description{}
\end{figure*}

The refinement workflow begins once an initial video has been generated. Instead of redrawing or modifying the entire sketch storyboard to produce a new result, users can refine the output directly within the video itself. As shown in Fig.~\ref{fig:refinement_cue}, the interface processes the video into a set of keyframe images, which act as semantic anchors that provide stable, context-rich entry points for targeted edits. 
Users can select a target keyframe in which they want to modify the related animation, and express refinement intention in two complementary ways. They may sketch directly on the paused frame to indicate spatial or motion adjustments (e.g., extending a trajectory line to lengthen an object’s movement). Alternatively, they may issue concise textual refinements such as \q{fade slower} or \q{loop twice.} 
To enable refinement, we implement two key steps.
By combining the immediacy of visual sketching with the precision of lightweight textual input, this approach supports a hybrid refinement workflow that flexibly captures both spatial and temporal aspects of animation.

To enable such refinement interaction, we implement two key steps. First, to provide interpretable keyframes as anchors for users to apply intentions, we use the VLM to analyze the initial animation code, detect salient timestamps, and expose them as keyframes. Second, to provide full context for localized video editing, we supply the VLM with user refinements (sketched on a keyframe or expressed in text), the timestamp of the keyframe, all extracted keyframes, and the initial animation to be modified. The VLM then updates only the relevant portion of the code, and the revised code is rendered into an updated preview, ensuring efficient and focused iteration.

\subsection{Study Design}
We recruited eight new participants (P1–P8; four male, four female) from the same university, with an average age of 25.8 years (SD = 3.7). None had participated in the previous stages. At this stage, participants were given access to the full set of mechanisms introduced in the paper, including both clarification cues and refinement cues.

The task design followed the same open-ended format as in earlier stages: participants were asked to create a short explainer-style animation of a concept of their choice using free-form sketching. However, to emphasize iterative creation and refinement, each participant was instructed to generate an initial video and then improve it through at least one round of refinement using the new mechanism. Since the refinement process required more time than earlier tasks, participants were asked to complete one animation project within the 30-minute session.

\begin{figure*}[ht!]
  \centering 
  \includegraphics[width=0.95\linewidth]{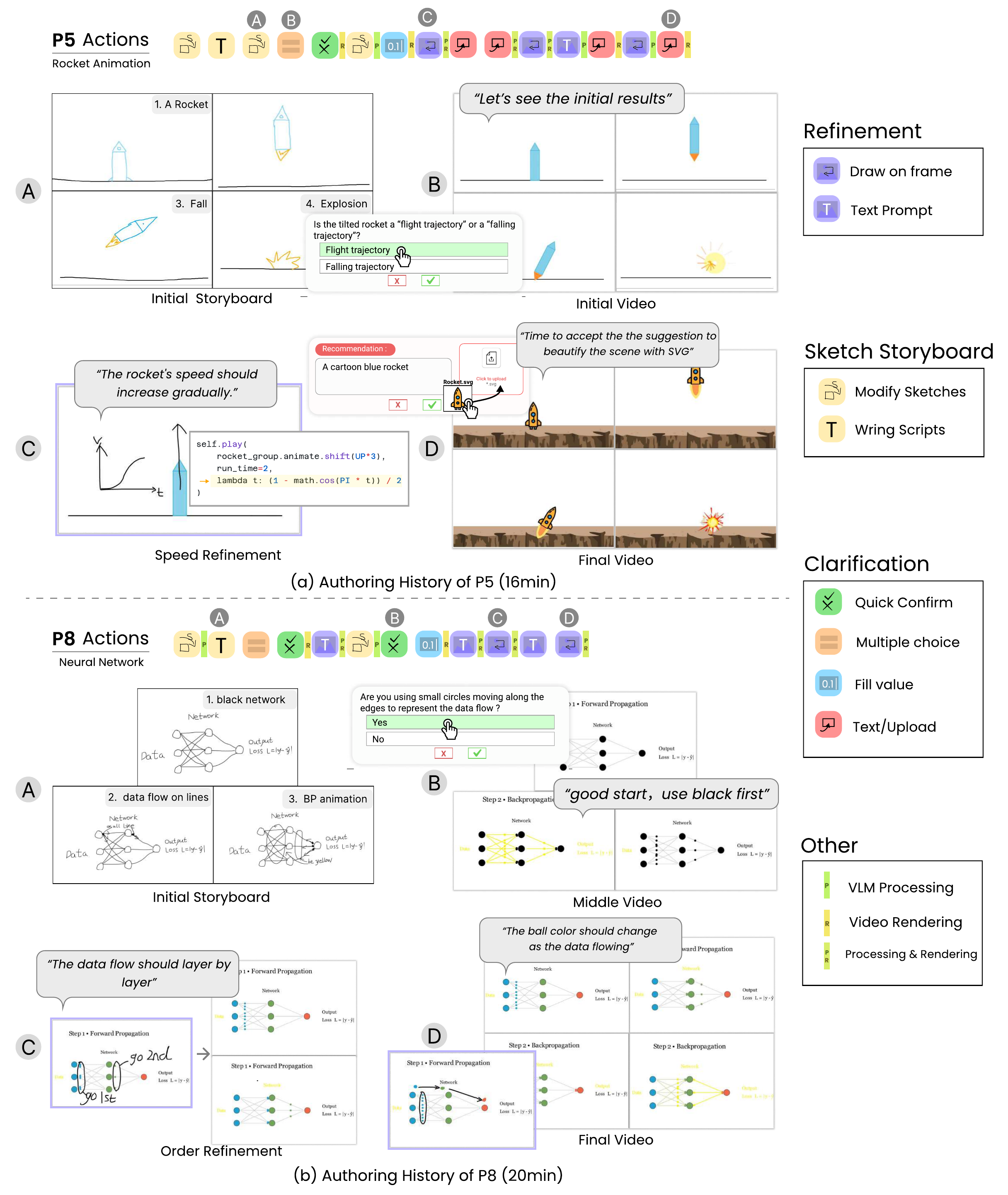}
  \caption{%
    Authoring process of two participants. The actions are sequentially listed (only resolved clarification cues are listed), and representative steps are selected for illustration.
  }
 
  \label{fig:authoring_process}
  \Description{}
\end{figure*}

\subsection{Results}
\subsubsection*{\textbf{Overall Use of Refinement Cues}}
Across the eight participants, we collected 12 edited videos with a total of 55 refinement operations. On average, participants performed 4.6 refinements per task, though the distribution was skewed: some participants stopped after one or two small adjustments (P2, P5), while others explored more iterative editing cycles with up to eight refinements (P7). Most refinements (36/55) were sketch-based, while the others were text-based. This division reflected a functional split: spatial animations (e.g., paths, shape sizes, directions) were usually expressed visually, while temporal refinements (e.g., timing, repetition) were handled with short text cues.

\subsubsection*{\textbf{Efficiency and Locality}}
Participants emphasized the efficiency of directly correcting generated results instead of redrawing entire storyboards. In 10 of the 12 final outputs, users reported that the unaffected portions of the video remained stable, which they considered crucial to maintaining creative momentum. 
\revise{In the user ratings (Fig.~\ref{fig:rating}), although completing animations in this stage took longer than in the previous stages, participants did not report a corresponding increase in perceived effort. This suggests that the interface interaction is efficient even with the added steps, and the improved animation outcomes justified the additional time.}
In interviews, six participants noted that this locality reduced frustration: \q{I don’t have to start over again—just fix the part I don’t like} (P4). Compared to Stage 1, where small sketch changes could lead to unrelated alterations, Stage 3’s refinement preserved coherence and increased trust in the system.

\subsubsection*{\textbf{Preferred Strategies}}
Different editing strategies were observed. Some participants (P3, P5, P7) intervened early, pausing the video within the first few seconds to adjust motions before they propagated. Others (P1, P2, P8) preferred to watch a complete draft first and then make targeted corrections. These choices also influenced how edits were sequenced. For instance, as shown in Fig.~\ref{fig:authoring_process}, P8 ensured that the overall structure of the neural network animation was correct before refining details such as layer-wise flow and color changes, describing the process as \q{just small fixes once it was understood.} In contrast, P5 followed a step-by-step approach, editing the rocket’s trajectory and fall before addressing the explosion and pacing, noting a preference to go \q{scene by scene.} Strategy selection was shaped not only by personal preference but also by the nature of the task and the system’s prior knowledge. In P8’s case, the VLM correctly recognized the rough sketch as a fully connected network and applied concepts such as backpropagation, which made the creation process more automatic and reduced the need for low-level specification. As P8 remarked, it felt like the system \q{already knows what I mean,} which shaped his choice to focus on high-level adjustments rather than redrawing details.

\subsubsection*{\textbf{User Perception of Control}}
Seven out of eight participants reported that refinement cues gave them a stronger sense of control than in previous stages. They valued being able to \q{tell the AI exactly what’s wrong, instead of hoping it guesses right} (P7). This contributed to a perceived shift from one-shot generation to an iterative, collaborative process. However, two participants also tested the system with higher-level semantic requests (e.g., \q{make this look more natural}), which were not successfully interpreted. These limitations revealed boundaries of the refinement mechanism: while effective for concrete spatial-temporal edits, it struggled with abstract stylistic or semantic adjustments. Nonetheless, participants widely agreed that refinement cues transformed the interaction into a more predictable and empowering authoring experience.

\section{Generalizability and Extensibility}
Beyond motion graphics, our concept can be extended to other forms of sketch-based dynamic content. Here we present a design outlook that builds on our prototype to envision alternative styles of output. While we have not implemented a complete system for these scenarios, we illustrate how the proposed pipeline, which combines vision–language model interpretation with clarification and refinement cues, can be organically integrated with other AI models to explore broader possibilities for sketch-driven creation.

\begin{figure*}[h!]
  \centering 
  \includegraphics[width=1\linewidth]{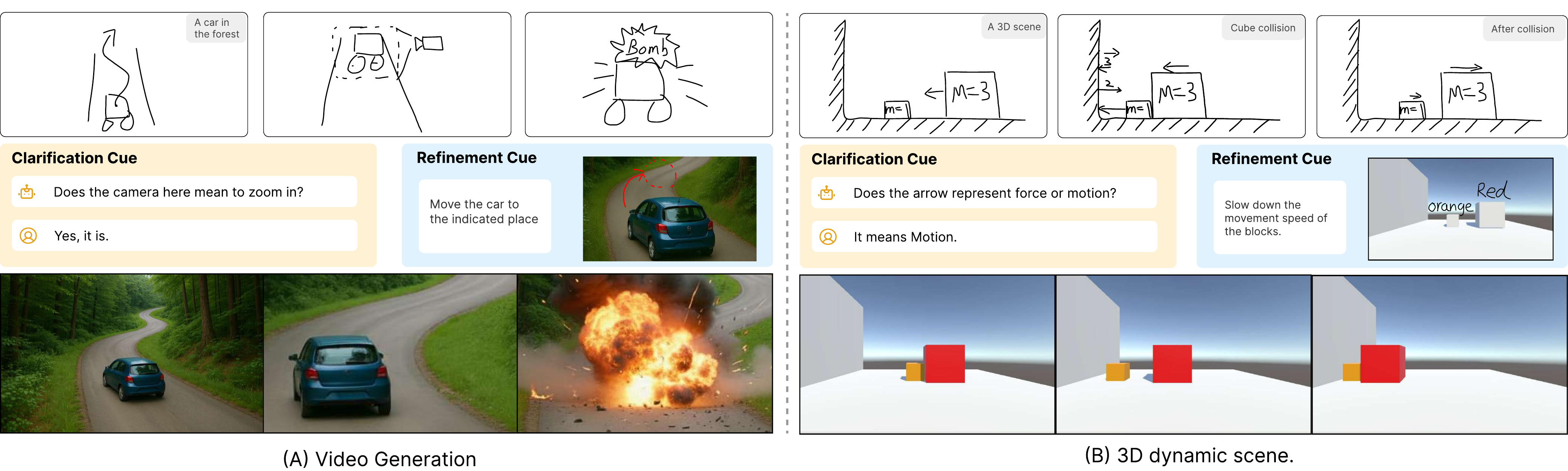}
  \caption{%
  Demonstration of extensibility of \tool. (A) Video generation example of a car driving along a forest road and exploding. (B) 3D dynamic scene example of block collision simulation. Top: user-drawn sketch storyboard. Middle: interactive clarification and refinement cues. Bottom: corresponding output video frames.
  }
 
  \label{fig:DesignOutlook}
  \Description{}
\end{figure*}

\subsection{Video Generation}
Currently, video generation has garnered significant attention~\cite{videocraft2025Li}, with recent work such as SketchVideo~\cite{liu2025sketchvideo} leveraging sketch storyboards to produce videos. However, that approach primarily relies on detailed keyframe layouts, which demand considerable drawing expertise and are often accompanied by extensive textual scripts. In contrast, our approach emphasizes minimal sketches, accessible to a broad range of users, for expressing abstract concepts, thereby lowering the entry barrier. 
Through the integration of these sketches with our proposed mechanisms, we enrich their interpretive depth and provide stronger support for video generation. At this stage, our demonstration is still conceptual and involves partial manual implementation: we primarily use our prototype to generate prompts for external image and video generation models, with the core contribution lying in the translation of user intent into effective prompts (see Fig.~\ref{fig:DesignOutlook}-A).
\revise{Compared with commercial systems such as Higgsfield’s ``draw-to-video'' feature\footnote{\url{https://higgsfield.ai/flow/draw}}
, which generates videos by annotating over existing images, our concept enables users to communicate their intent before generation through iterative, back-and-forth interaction, even when the initial input is an abstract sketch drawn from scratch.}

\subsection{Dynamic 3D Scene }
Beyond 2D motion graphics, 3D dynamic scenes are widely applied in domains such as games and explanatory videos~\cite{anicraft2024li}. To illustrate this extensibility, we extended our concept to 3D animation within Unity. As shown in Fig.~\ref{fig:DesignOutlook}-B, the process begins with a sketch storyboard in which the user draws two cubes and a wall, using arrows to indicate motion and collision. This example demonstrates how the proposed mechanisms can assist the generation process by supplementing sketches with detailed attributes. In practice, we envision integrating our pipeline with Unity’s Model Context Protocol (MCP), enabling the VLM to interact directly with the Unity Editor by issuing function calls or generating C\# scripts after interpreting the sketch. Such a workflow would allow storyboard concepts to be rapidly transformed into executable 3D scenes.

\section{Discussion}
We began with the question of how humans can communicate dynamic intent to AI through free-form sketches. Across three stages, we progressively shaped this process through our proposed design interventions, first by probing expressiveness, then by supporting clarification, and finally by enabling refinement. This revealed both the potential and the limits of sketch-based interaction. In the discussion, we connect these results to broader themes and outline implications for future research.

\subsection{Synergistic Cue Integration}
\revise{Our integrated system reveals that clarification and refinement are not merely sequential steps, but complementary mechanisms that support the progressive formalization of intent. We observe a distinct division of labor: Clarification Cues (Stage 2) act as a semantic guardrail before generation, resolving high-level structural ambiguities (e.g., distinguishing a trajectory path from a shape outline) to ensure a correct foundation. In contrast, Refinement Cues (Stage 3) address perceptual imprecisions after generation, enabling users to tune local sensory details (e.g., timing or color) that are often difficult to articulate upfront. This "disambiguate-then-refine" synergy enables a paradigm of "lazy specification": users can start with low-effort, abstract sketches, relying on the system to proactively align high-level intent while deferring precise adjustments to the post-generation phase. This shifts animation authoring from a high-stakes, one-shot task to a fluid, collaborative dialogue where intent is shaped iteratively.} 

\subsection{Shaping Intent with AI}

In our study, we observed that users often do not have a fully-formed animation intent at the beginning of the authoring process. Instead, their intent gradually evolves through interaction and iteration. Similar observations have been made in other domains involving generative AI~\cite{Lee2025When}, where the core challenge lies in supporting users in shaping their ideas fluidly rather than expecting them to articulate a complete intent upfront. While some existing systems attempt to address this by increasing output diversity, they often lack precise alignment with user intentions due to limited vision-language understanding capabilities. Other systems rely on complex, multi-step prompt engineering workflows to help users refine their ideas, but these often interrupt the creative flow.

We argue that sketching provides a lightweight and effective medium for expressing and gradually refining animation intent. Prior systems, such as K-Sketch~\cite{davis2008ksketch}, have highlighted that traditional animation tools are often too complex for novices due to their emphasis on frame-by-frame control. In contrast, simple sketches can allow users to communicate temporal dynamics without extensive effort, making the process significantly more accessible.
In our exploration, users can either draw rough sketches from scratch or annotate on top of generated results. This supports a more fluid, back-and-forth interaction with the VLM understanding and generation. For example, during Stage 3 of our study, users often began sketching without a clear goal in mind. It was only when presented with clarification cues—such as candidate suggestions or disambiguation options—that they began to consciously shape their intent. This observation echoes findings from recent work like AnyAni~\cite{qiu2025anyaniinteractivegenerativeai}, where novice users struggled to articulate animation needs and required iterative refinement through system feedback.

Crucially, we avoid a traditional end-to-end generation pipeline in which the AI produces a complete result in one shot, leaving little room for user intervention. Instead, we preserve the opportunity for users to continually influence and reinterpret how the AI understands their sketches. This not only allows users to refine the output but also helps them reflect on and reshape their own ideas. Through this iterative, co-creative loop, the final animation gradually emerges—balancing the strengths of both human intuition and AI generation.


\subsection{Sketching Strategies}

Our study reveals that participants naturally employed sketches at different levels of abstraction. On the one hand, \textit{high-level sketches} were frequently used to represent rough shapes, motion directions, or symbolic marks. On the other hand, users occasionally turned to \textit{low-level sketches}, such as precise motion paths or data diagrams, when finer control was needed.
Because our system relies primarily on the reasoning capabilities of VLMs, it was more effective at interpreting high-level sketches that conveyed semantic intent, while its ability to handle low-level details was less consistent. In practice, this meant that abstract annotations were readily understood, substantially lowering the entry barrier for novice users~\cite{davis2004informal,davis2008ksketch}, whereas capturing accurate shapes or trajectories often depended on chance alignment with the model’s visual interpretation.
More generally, the two levels demand different forms of technical support. Low-level sketches require specialized algorithms to preserve geometric fidelity and temporal precision, as in ControlNet~\cite{zhang2023controlnet}, whereas high-level semantics benefit more from the broad reasoning capacities of VLMs~\cite{liu2025logomotion}. 
Future work should explore hybrid authoring workflows that fluidly integrate both modalities. Adaptive systems that allow users to begin with high-level, accessible annotations and selectively refine with low-level precision sketches could enable smooth transitions between rapid ideation and detailed specification, aligning authoring effort with creative goals.

\revise{In our analysis, we found participants used two main ways to sketch dynamic intent.
The first was in-image motion annotations, where they drew arrows, motion curves, or ghosted outlines directly onto a single frame. This was a concise way to define a specific trajectory or transformation. The second was across-frame sequencing, the classic storyboard method where they conveyed motion by drawing successive keyframes. This approach naturally encoded temporal evolution and was better for showing complex compositional changes.
This duality reflects a clear trade-off between efficiency and precision. In-image annotations were fast and preferred for simple translations or rotations. Sequencing, while more work, offered far more control over timing and complex interactions.
Interestingly, each mode created unique problems for the VLM. A frame crowded with in-image annotations could become a mess of conflicting cues. Yet with sequencing, if the temporal gap between two frames was too large, the model might misinterpret a single moving element as two completely separate objects.}

\subsection{Sketch Inherent Uncertainty}

VLMs demonstrate impressive multimodal reasoning capabilities, yet they still struggle to interpret sketches in ways that align with creator intent—especially when users~\cite{mukherjee2023SEVA}, particularly novices, are limited to quick, abstract drawings due to time constraints~\cite{lin2020it}. This becomes a primary source of ambiguity in sketch-based inputs. In the context of dynamic content creation, this challenge is further amplified: static sketches are often underspecified for expressing animation, lacking temporal information, clear motion styles, or distinctions between annotations and scene elements. These limitations make animation sketches uniquely ambiguous.

While the model’s ability to detect such ambiguity inevitably depends on its reasoning capacity and the quality of prompt engineering, our findings suggest that ambiguity should not be viewed merely as a technical flaw, but rather as a generative opportunity. In our Stage 2 study, many ambiguities surfaced by the VLM aligned with users’ own intuitions, helping reveal aspects of intent that were not fully formed. In other cases, mismatches between model and user interpretations prompted valuable reflection and clarification. There were also instances where the model failed to detect ambiguity that seemed obvious to the user. Rather than dismissing these discrepancies as failure, we frame them as moments of negotiation—opportunities for users to refine their intent and guide the model’s interpretation. This reflects a shift from deterministic automation to an interactive, human-in-the-loop workflow. Importantly, we did not fine-tune a task-specific VLM, but instead explored how general-purpose models might already support this kind of collaborative meaning-making. Future work could extend this foundation by incorporating sketch-specific datasets or animation-aware reasoning modules to improve the alignment between model inference and user intent.

\subsection{Sketch Parsing}
In this study, we employed the powerful visual understanding capabilities of VLMs to parse users’ free-hand sketches. However, due to the limitations of current general-purpose VLMs, sketches had to be exported as images and analyzed at the pixel level, rather than being further parsed in the vector space. As a result, the VLM could provide an overall semantic interpretation, but was unable to fully leverage the structural and temporal information embedded in stroke data.
While prior work has explored stroke-level analysis~\cite{wobbrock2007gestures}, existing methods are limited to basic classification and predefined symbols, which undermines support for true free-form authoring.
There are also attempts that have explored feeding SVGs or stroke sequences into LLMs for interpretation, but purely textual reasoning often struggles to capture 2D spatial information~\cite{Vinker_2025_CVPR}. Consequently, our system could not effectively explain the specific meaning of a stroke(e.g., strokes that function as static layouts, dynamic annotations). In free-hand drawing scenarios, this complexity became even more pronounced, as the stroke may play multiple roles and can not be easily classified. 
Our findings point toward the need for a cross-modal VLM specifically designed for sketch parsing in future work. Compared with pixel-only input, integrating vector- and stroke-level parsing could unlock richer opportunities, as stroke sequences data inherently encode order, timing, and grouping cues essential for animation authoring. For example, if a stroke is drawn at the end, it might be a animation anootation. This also open new possibilities for sketch-based authoring systems, where the systems can reason more naturally about the sketching process and provide responsive feedback.

\subsection{Limitations}
This work has several limitations. 
1) \textit{Participant diversity:} The participant pool was relatively homogeneous, consisting mainly of university students with some background in graphics or video, and thus may not fully capture perspectives from children, complete novices, or professional animators. 
2) \textit{Real-time interaction:} The system does not yet provide instant feedback at the stroke level, as sketches need to be translated into code and rendered into video, which typically takes several seconds. 
3) \textit{Model dependency:} The approach currently builds on general-purpose VLMs, whose performance is shaped by their training data and reasoning mechanisms, and future advances in these models may influence the outcomes. 
4) \textit{Task scope:} The study focused on short storyboards and motion graphics (tens of seconds), given our emphasis on sketching and the higher cost of generating longer videos. Extending the approach to longer or more complex animations is a promising avenue for future work.
\revise{5) \textit{Expressive range:} The current implementation is constrained by the capabilities of the Manim library, which primarily supports procedural 2D vector animations and simple transformations (e.g., translation, rotation, scaling, and color changes). It does not yet accommodate more expressive animation forms such as character motion, fluid deformations, or physics-based dynamics.}

\section{Conclusion}

This paper explores how free-form sketches can serve as an intuitive medium for expressing animation intent in the automatic creation of dynamic content. Using simple motion graphic scenarios as a starting point, we investigate the alignment between user intentions and generated results. Our overarching goal is to enable novice users to communicate animation ideas through minimal, informal inputs and to have the machine effectively interpret and realize these intentions.
We improve the process of simple static sketches to convey dynamic intent across three stages: 
In the first stage, we support a basic interface to transform free-form sketch storyboards to motion graphics with the assistance of VLM. Observing the ambiguity and misalignment in this initial interaction, the second stage introduces clarification mechanisms that surface and resolve uncertainty in the sketch interpretation process. Building on this clarified understanding, the third stage incorporates intuitive visual post-editing tools, enabling users to close the gap between user intention and the generated outcome.
Together, these stages form a continuous feedback loop that transforms VLMs from one-shot generators into interactive collaborators, capable of handling uncertainty, supporting intent refinement, and empowering users to co-construct dynamic content through sketch-based interaction.

\begin{acks}

This work was partially supported by the National Natural Science Foundation of China (Project No. 62502410).

\end{acks}

\bibliographystyle{ACM-Reference-Format}
\bibliography{reference}

\clearpage
\appendix
\begin{figure*}[ht!]
    \centering
    \includegraphics[width=0.8\linewidth]{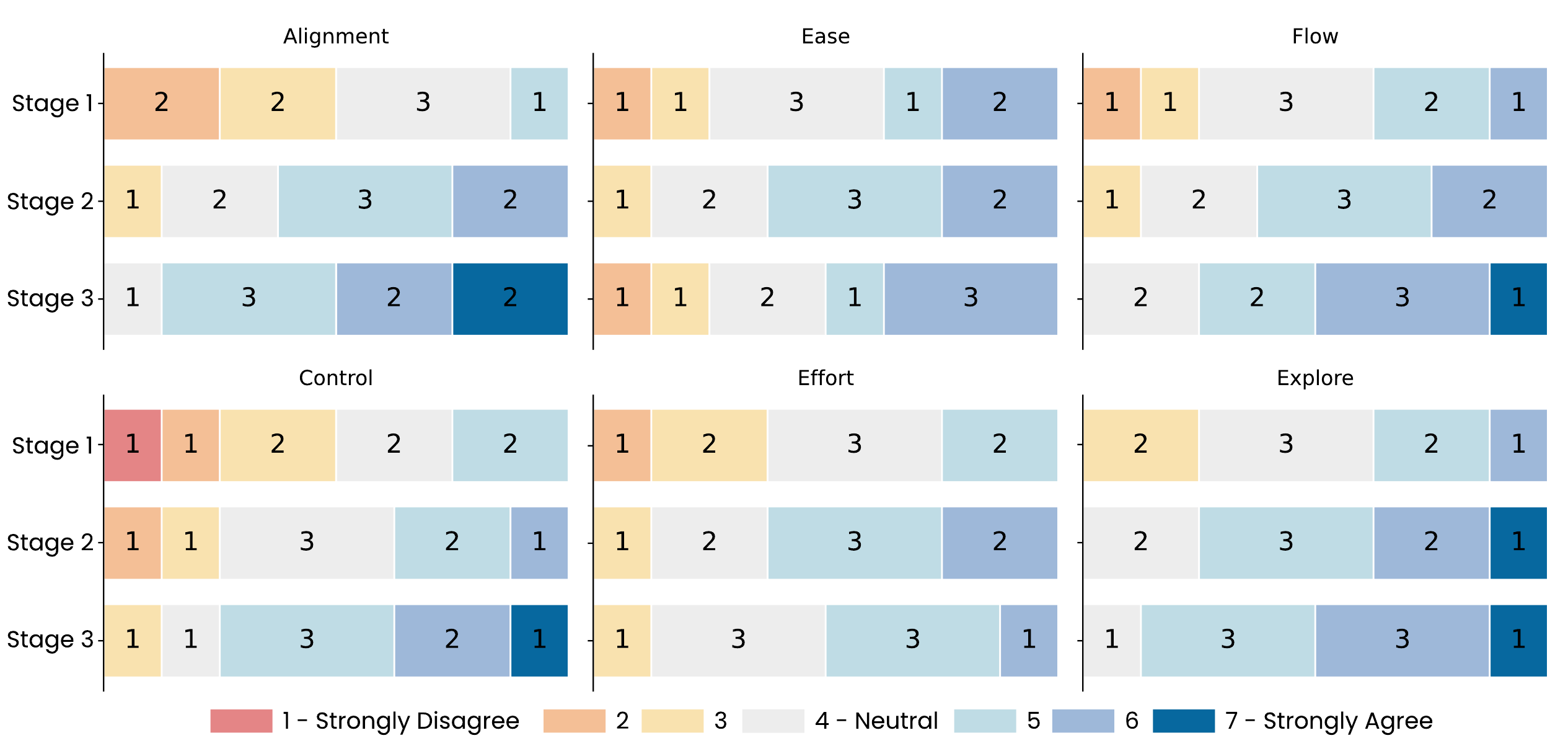}
    \caption{Self-defined Likert scale results across the three stages. Alignment: The generated results reflected what I had in mind. Ease: The interface was easy to use and understand. Flow: The workflow felt smooth without disrupting my creative process. Control: I felt in control of how my input was interpreted. Effort: The effort required to use the system was reasonable. Explore: The system encouraged me to try new ways of expressing my ideas.}
    \label{fig:rating}
    \Description{}
\end{figure*}
\section{User Rating in Three Stages}
\revise{Fig.~\ref{fig:rating} illustrates the cumulative benefit of our design interventions across the three stages. We observe a substantial upward trend in Alignment and Control, shifting from predominantly negative ratings in Stage 1 to highly positive ones in Stage 3. This validates that while the VLM alone (Stage 1) often struggles with interpretation, the integration of clarification and refinement effectively bridges the gulf between user intent and system output. Notably, despite the additional interaction steps introduced in the final system, Effort ratings improved compared to Stage 1. This suggests that users perceived the targeted refinement process as significantly less laborious and frustrating than the repetitive trial-and-error redrawing required in the baseline. Furthermore, Flow and Ease remained consistently high, indicating that the added mechanisms were seamlessly integrated without disrupting the creative momentum.}

\section{Visual Understanding Capability Experiment}
Before the main study, we conducted pilot tests to check feasibility, focusing on prompt design and the VLM’s ability to interpret free-form sketches. The results were encouraging: the prompts guided the model effectively, and the VLM showed strong potential in handling abstract input.

Here we show how the VLM with prompt engineering interpreted several Stage 1 sketches, illustrating how rough or symbolic drawings were mapped into animation concepts. This is not part of our core contribution but serves as an important foundation for the paper.

\revisemb{
\section{User Study Survey and Interview Questions}
\label{sec:appendix-survey}

This appendix provides the detailed instruments used in our three-stage user study to evaluate \textit{SketchDynamics}. These include the participant background survey, the post-task Likert scale questionnaire, and the semi-structured interview guide.

\subsection{Pre-study Questionnaire: Background and Experience}
Before the experiment, participants provided background information to help us understand their prior expertise in animation and sketching:
\begin{itemize}
    \item \textbf{Demographics}: Age, Gender, and Handedness.
    \item \textbf{Motion Graphics Experience}:
    \begin{itemize}
        \item How often do you watch motion graphics or explainer videos? (1: Never, 7: Daily).
        \item Have you ever created motion graphics using tools such as After Effects, CapCut, or others?
        \item Rate your proficiency in video editing and animation software (1: Novice, 7: Expert).
    \end{itemize}
    \item \textbf{Sketching Experience}:
    \begin{itemize}
        \item How often do you use sketching for ideation or communication? (1: Never, 7: Daily).
        \item Rate your comfort level with drawing/sketching (1: Very Uncomfortable, 7: Very Comfortable).
    \end{itemize}
\end{itemize}

\subsection{ Post-task Questionnaire}
After completing the tasks in each stage, participants rated their experience based on the following statements using a 7-point Likert scale (1: Strongly Disagree, 7: Strongly Agree)

\begin{itemize}
    \item \textbf{Alignment}: The generated results reflected what I had in mind.
    \item \textbf{Ease}: The interface was easy to use and understand.
    \item \textbf{Flow}: The workflow felt smooth without disrupting my creative process.
    \item \textbf{Control}: I felt in control of how my input was interpreted.
    \item \textbf{Effort}: The effort required to use the system was reasonable.
    \item \textbf{Explore}: The system encouraged me to try new ways of expressing my ideas.
\end{itemize}

\subsection{ Semi-structured Interview Guide}
The following questions guided our qualitative analysis across the three stages of the study:

\subsubsection{General Questions (All Stages)}
\begin{itemize}
    \item What was your primary strategy for translating animation ideas into free-form sketches? 
    \item Were there specific animation intents that you found particularly difficult to express through sketching?
    \item Did the system's interpretation of your sketches surprise you? In what way? 
\end{itemize}

\subsubsection{Stage 2: Clarification Cues}
\begin{itemize}
    \item How did you perceive the clarification requests (e.g., multiple-choice, fill-value)? Did they feel like interruptions or helpful checks? 
    \item Did the clarification process help you better define animation details (like timing or speed) that you hadn't initially considered?.
\end{itemize}

\subsubsection{Stage 3: Refinement Cues}
\begin{itemize}
    \item How does the experience of refining a generated video compare to redrawing a storyboard from scratch?
    \item Did the ability to edit specific video frames increase your sense of control over the AI collaborator? 
    \item Which did you find more useful for refinement: drawing on the frame or providing text prompts? Why? 
\end{itemize}
}

\begin{table*}[ht!]
\centering
\caption{Examples of sketch types and the animation intentions they convey}
\begin{tabular}{p{0.12\textwidth} p{0.20\textwidth} p{0.15\textwidth} p{0.38\textwidth}}
\toprule
\textbf{Sketch Type} & \centering \textbf{Sketch Example} & \textbf{Text Script} & \textbf{Animation Description by VLM} \tabularnewline
\midrule

\raggedright Camera Motion  
& \centering $\vcenter{\hbox{\includegraphics[width=0.12\textwidth]{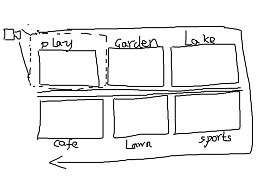}}}$
& \raggedright Virtual Camera Movement
& \raggedright Plan for a virtual camera sweep across six regions in a Z-shaped path: Play $\to$ Garden $\to$ Lake (top row), then down to Sports $\to$ Lawn $\to$ Cafe (bottom row), followed by an exit movement to the left. \tabularnewline \addlinespace

\raggedright Fade Out
& \centering $\vcenter{\hbox{\includegraphics[width=0.12\textwidth]{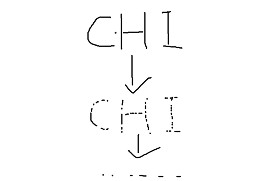}}}$
& \raggedright Null
& \raggedright The word ‘CHI’ appears, then gradually fades while moving downward (represented with dashed lines). \tabularnewline \addlinespace

\raggedright Shape Changing
& \centering $\vcenter{\hbox{\includegraphics[width=0.12\textwidth]{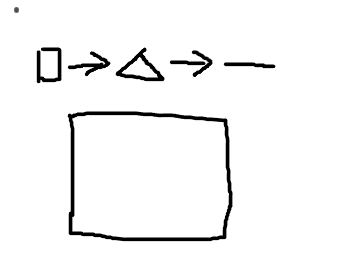}}}$
& \raggedright Null
& \raggedright A square morphs into a triangle and then into a line, shown sequentially within the canvas. \tabularnewline \addlinespace

\raggedright Moving Direction
& \centering $\vcenter{\hbox{\includegraphics[width=0.12\textwidth]{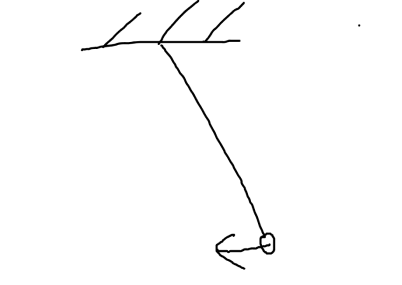}}}$
& \raggedright Null
& \raggedright A pendulum suspended from a pivot with an arrow pointing left at the bob, indicating oscillatory motion. \tabularnewline \addlinespace

\raggedright Ghost State
& \centering $\vcenter{\hbox{\includegraphics[width=0.12\textwidth]{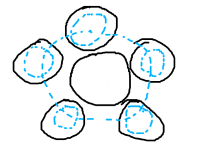}}}$
& \raggedright Blue indicates future
& \raggedright The central circle enlarges to the blue dashed size while the five surrounding circles shrink to their dashed sizes. \tabularnewline

\bottomrule
\end{tabular}
\end{table*}

\end{document}